\documentclass[%
superscriptaddress,
 amsmath,amssymb,
 aps,
pra,
twocolumn
]{revtex4-2}

\usepackage{graphicx}
\usepackage{dcolumn}
\usepackage{bm}
\usepackage{appendix}

\begin{document}


\title{Strain dependent viscous response describes the mechanics of cohesionless soft granular materials}
\author{Chandan Shakya}
\email{c.shakya@uva.nl}
\affiliation{Van der Waals-Zeeman Institute, Institute of Physics, University of Amsterdam, Amsterdam, The Netherlands.}
\affiliation{Physical Chemistry and Soft Matter, Wageningen University.}

\author{Luca Placidi}%
 \email{luca.placidi@uninettunouniversity.net}
\affiliation{%
 Faculty of Engineering, International Telematic University UNINETTUNO, CorsoVittorio Emanuele II 39, 00186 Rome, Italy
}%

\author{Anil Misra}
\email{anil.misra@fiu.edu}
\affiliation{
 Department of Civil and Environmental Engineering, Florida International University, Miami, U.S.A.
}%

\author{Lars Kool}
\email{lars.kool@espci.fr}
\affiliation{Laboratoire de Physique et M\'echanique des Milieux H\'et\'erog\`enes (PMMH), UMR 7636 CNRS, ESPCI Paris, PSL Research University, Sorbonne Universit\'e and Universit\'e Paris Cit\'e,
7-9 Quai Saint-Bernard, Paris, 75005, France}
\affiliation{Institut Pierre Gilles de Gennes (IPGG), ESPCI, Paris, France }

\author{Anke Lindner}
\email{anke.lindner@espci.fr}
\affiliation{Laboratoire de Physique et M\'echanique des Milieux H\'et\'erog\`enes (PMMH), UMR 7636 CNRS, ESPCI Paris, PSL Research University, Sorbonne Universit\'e and Universit\'e Paris Cit\'e,
7-9 Quai Saint-Bernard, Paris, 75005, France}

\author{Jasper van der Gucht}
\email{jasper.vandergucht@wur.nl}
\affiliation{Physical Chemistry and Soft Matter, Wageningen University.}

\author{Joshua A. Dijksman}
\email{j.a.dijksman@uva.nl}
\affiliation{Van der Waals-Zeeman Institute, Institute of Physics, University of Amsterdam, Amsterdam, The Netherlands.}
\affiliation{Physical Chemistry and Soft Matter, Wageningen University.}

\date{\today}

\begin{abstract}
Granular materials are ubiquitous in nature and are used extensively in daily life and in industry. The modeling of these materials remains challenging; therefore, finding models with acceptable predictive accuracy that at the same time also reflect the complexity of the granular dynamics is a central research theme in the field. Soft particle packings present additional modeling challenges, as it has become clear that soft particles also have particle-level relaxation timescales that affect the packing behavior. We construct a simple one-dimensional, one-timescale model that replicates much of the essence of compressed hydrogel packing mechanics. We verify the model performance against both 3D and 2D packings of hydrogel particles, under both controlled strain and stress deformation conditions. We find that the modification of a Standard Linear Solid model with a strain dependent prefactor for the relaxation captures the time-history and rate dependence, as well as the  necessary absence of cohesion effectively. We also indicate some directions of future improvement of the modeling.
\end{abstract}

\maketitle


\section{\label{intro}Introduction}

Granular materials are abundant and are present in all spheres of human activities in both nature and industrial processes.  Perhaps because they are so plentiful and because they have been around for so long, their surprising mechanical features are often overlooked. One of these features is the presence of slow relaxation effects. Slow relaxation effects manifest themselves in phenomena such as creep \cite{smith2015self,deshpande2023sensitivity,deshpande2021perpetual,srivastava2017slow}, relaxation \cite{eftekharnia2023understanding} and consolidation \cite{leonards1960time}, among others.  These effects are not well understood, but have been linked to a multitude of different microscopic phenomena, including thermal fluctuations and weathering \cite{jiang2015effect}, mechanical noise from shear \cite{agoritsas2016driven}, localized sporadic fluidization of granular materials \cite{deshpande2023sensitivity}, particle breakage \cite{lade1998experimental,mcdowell2003creep}, and periodic external driving \cite{deshpande2023sensitivity}. In packings made of softer grains, the time and rate dependent response of grain material adds to the complexity of the packing response \cite{osthoff1954chemical,islam2021poroelastic} which may even lead to the commonly observed loading rate dependence. Clearly, for granular materials, there are many different possible sources of relaxation, making it important to conceive simple models that capture the essence of the phenomenology exhibited at the macroscale. In this paper, we aim to describe recent observations on the mechanical response of soft, dissipative particle packings.

Many phenomenological models have been presented to account for these time and rate dependent changes. These models often include viscoelastic components such as the Kelvin-Voigt, the Maxwell or the Standard Linear model. However, such simple models fail to capture the slow time dependence of granular packings, and in particular the response of cyclic loading. The response to cyclic loading is challenging to capture, as it needs to incorporate the dissipative force loops, typically encountered in such experiments, while also satisfying the no-cohesion property typical for many granular materials: that is upon unloading to the reference state (or at incipient release of contact with the boundary of the material), any driving wall must be at zero stress. All standard viscoelastic models result in a tensile force when the material is decompressed to near zero strain in finite time. It is also worthwhile to note that granular assemblies can have different kind of dissipation mechanisms, which can be of rate independent type (such as those related to damage and plastic aspects as exemplified by the critical state phenomena) \cite{yilmaz2024emergence,ciallella2022shear} or of rate dependent type \cite{ciallella2022continuum,misra2014nonlinear,misra2015thermomechanics}. Constitutive equations that are non-local in time are also known in the literature often modeled using fractional-order calculus \cite{sun2017constitutive,assante2015higher}. The purpose of this paper is to consider a non-local in time constitutive prescription that is at the same time easier and valid for the experimental characterization of the considered samples. To this end we consider a modified standard Linear Solid model that captures these essential components in the cyclic loading response of dissipative, rate dependent granular systems. We are inspired to construct such models by experiments conducted on hydrogel packings. The cyclic loading experiments were done on both three dimensional (3D) and quasi two dimensional (2D) packings of hydrogel particles, at both controlled strain and stress conditions. 

Many details of the 2D and 3D experiments using particle packings tested in the paper can be found in \cite{shakya2024viscoelastic,timescale-inprep}. It is notable that the particles in these experiments are cohesionless with a relative density close to that of the medium (water) making gravitational forces negligible. In addition, the intrinsic relaxation timescales of the particles are available from independent measurements, which provides a way to compare the packing relaxation timescales with that of a single particle. Our previous work has focused on measuring the introsic dynamics of the hydrogel material~\cite{timescale-inprep}, revealing the existence of multiple timescales. It is also notable, that 3D experiments were displacement (strain) controlled. Further, the 2D experiments were performed using nominally circular disk-type cohesionless particles. These experiments were performed using stress controlled boundary conditions. A suite of cyclic experiments with different loading rates, displacement amplitudes, and wait times have been performed. We will show that the introduced modified standard Linear Solid model is able to replicate many of the interesting experimental observations, particularly those related to the effect of loading history, the lack of particle cohesion and loading rates.

\section{\label{setup} Strain control: 3D hydrogel sphere packings}
The setup used for experiments was inspired from the compression setup described by Brodu et. al.\cite{brodu2015spanning}. The setup consists of a rate-controlled movement system and a sensing system that operate independently of each other.  Movement and driving force measurements were achieved using a rod connected to an adjustable stage being driven by an actuator(M-451 High-Load Precision Z-Stage). The actuator is controlled by a motor controller (C-863 Mercury Servo Controller) that communicates with a computer via a Labview interface with drivers from Physik Instrumente (PI). 

\begin{figure}
    \centering
    \includegraphics[scale=0.2]{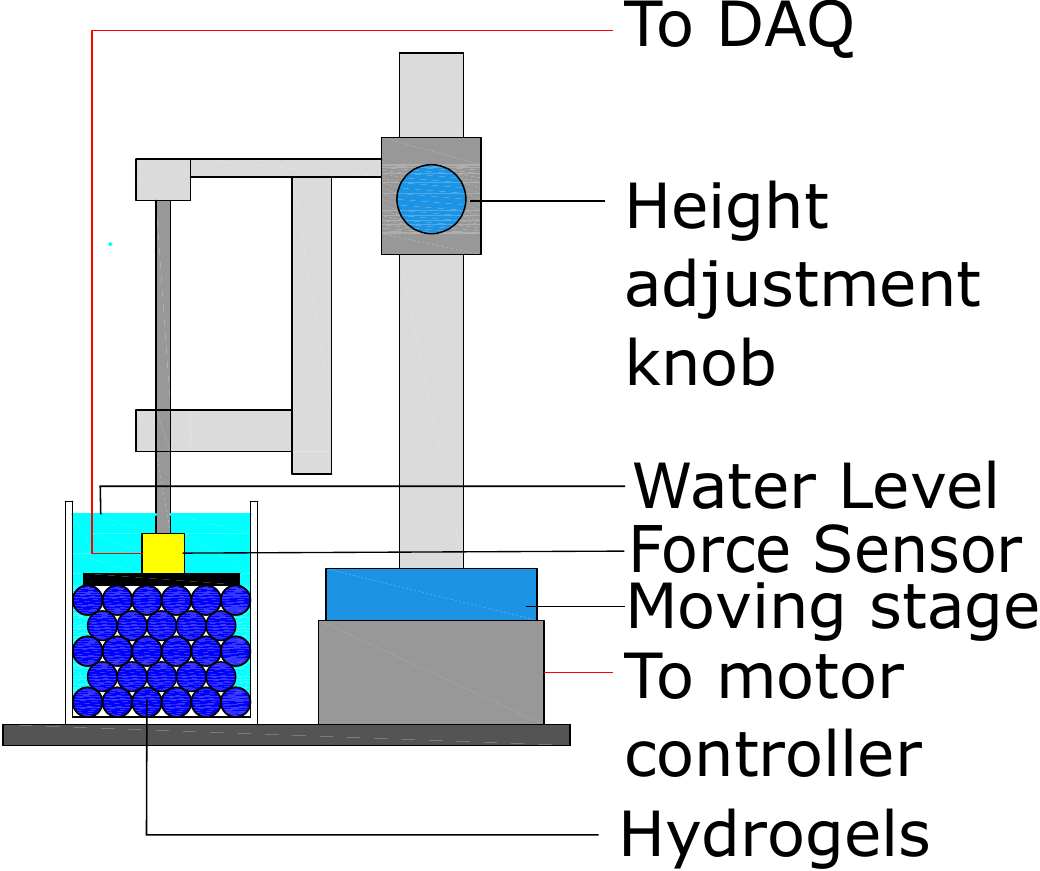}
    \includegraphics[scale=0.15]{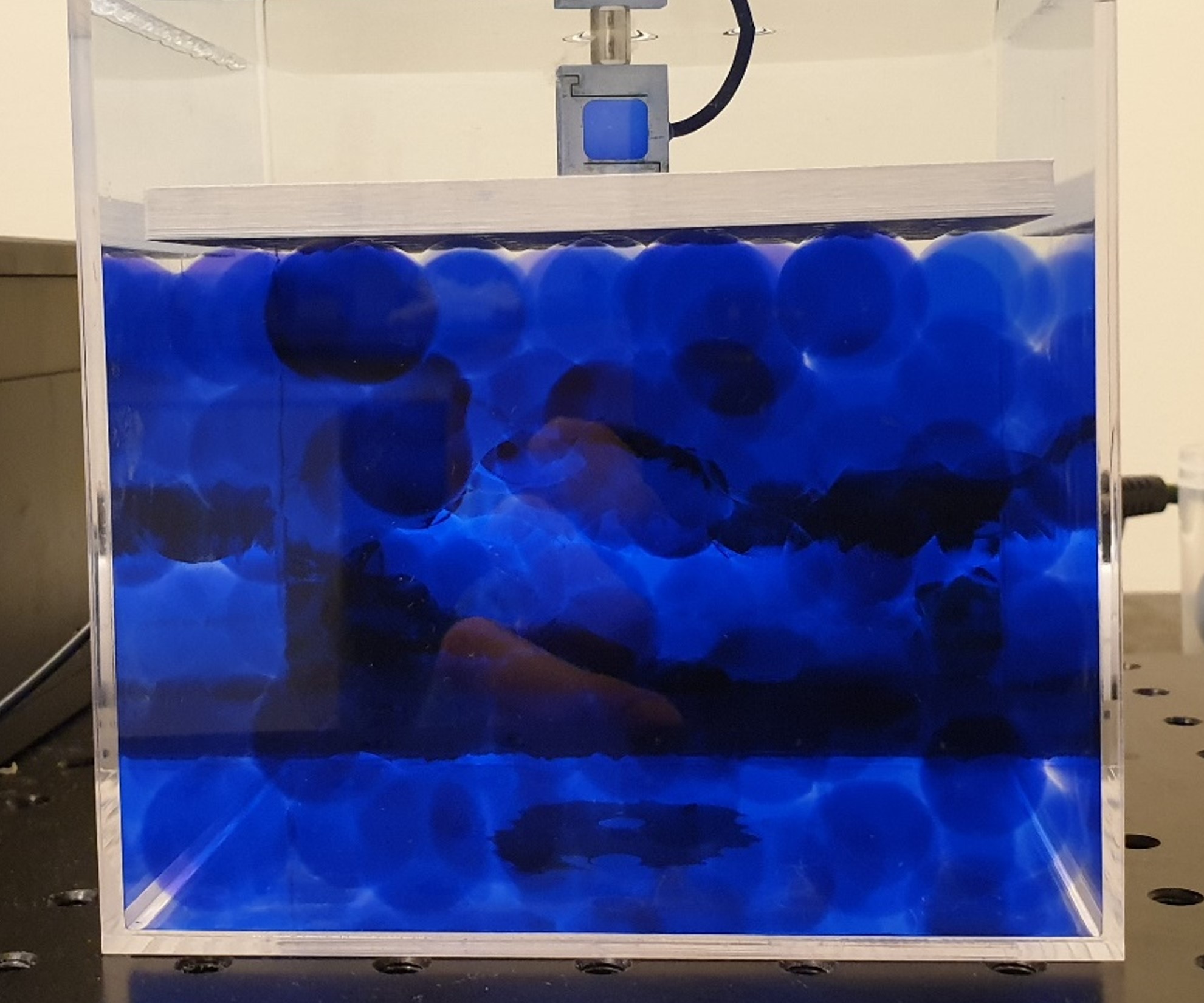}
    \caption{a) Schematic of the experimental setup b) picture of the prepared sample}
    \label{fig:setup}
\end{figure}

The rod applying the force was connected to an S-beam load cell (Futek LSB210 FSH03941) based on a Wheatstone bridge. Both the 10V input voltage to the load sensor and the outputs of the load cell are connected to a strain gauge amplifier(Futek CSG 110). The output from the amplifier is sent through a low-pass filter to a Data Acquisition Unit  (NI DAQ USB 6211) that converts the analog signals into digital signals that can be read by the computer. The force sensor is directly connected to a 3D  printed compression plate, which allowed for approximately 3 mm clearance an all sides with the acrylic box holding the hydrogels and had holes of 4 mm diameter holes in it to allow for easy flow of water. A schematic of the setup is shown in Fig.~\ref{fig:setup}a.

\subsection{\label{sample_preparation}Sample preparation}
The particles used are Polyacrylamide hydrogels manufactured by Educational Innovation Ltd. For these sets of experiments the dehydrated hydrogels were re-hydrated in milli Q water at laboratory temperature. The rehydrating milli-Q water was dosed with a small volume of Nile blue perchlorate dissolved in absolute ethanol to ensure an even distribution of the dye. The typical size of the hydrogels sphere was 18.5mm. The fully grown particles are then counted and placed in the rectangular acrylic boxes filled with milliQ water for testing.  

\subsection{\label{protocol}Initialization of the experiment}
The compression plate is slowly lowered into the rectangular acrylic boxes filled with hydrogels and water until the force sensor is fully submersed in water. In all the experiments the water level was approximately 4 mm above the top surface of the force sensor, ensuring that the force sensor remained underwater throughout the entire experiment and that any evaporation of water would not affect the force readings. The compression plate was lowered very slowly to ensure that no air bubbles are trapped when the compression plate comes in contact with the water layer. If air bubbles are visible underneath the compression plate, the plate had to be removed, tilted and lowered again until no air bubbles are trapped under it. A slight tilt during preparation was provided to the compression plate by loosening the height control knob in the setup. At this point, we ensured that the compression plate and force sensor were completely underwater, but the compression plate does not make contact with the hydrogels below it. The hydrogels and  the compression plate are separated by approximately 2-3 mm. The distance from the bottom of the acrylic box to the bottom of the compression plate is measured by sliding a vernier caliper in the clearance between the compression plate and the acrylic box ($l_{init}$). Once this distance was measured, the acrylic box was re-centered to make sure there is no interference between the compression plate and the walls of the acrylic box.

We used $l_{init}$ as a rough estimate of the packing size and calculate a rough compression velocity using a $\dot\epsilon$ of $2\times10^{-3}$ [/s] and indent the packing by 6-7 mm, then waited for 10 seconds and return the compression plate to its original position. This was done in 3 cycles to find a consistent point  of contact. Then the packing was allowed to relax for at least 1 hour. The compression plate was then placed at the determined point of contact and the precise packing height ascertained. Once the packing size had been determined, the packing could  be compressed at different velocities corresponding to different strain rates ($\dot\epsilon$) to different strains ($\epsilon$). The prepared sample can be seen in Fig.~\ref{fig:setup}b.

\subsection{\label{exp_protocols}Strain controlled experimental protocols}
In this section, we describe the various experimental protocols and applied strain profiles the packing was subjected to and the force responses obtained during the application of the different strain profiles. Table \ref{parameter_experiment} in Appendix \ref{para_meters} summarizes some of the experimental parameters used to apply load in the experimental protocols that will be discussed below. 

\subsubsection*{\label{complete_compression} Protocol 1: Relaxation test}
\begin{figure}
    \centering
    \includegraphics[scale=0.32]{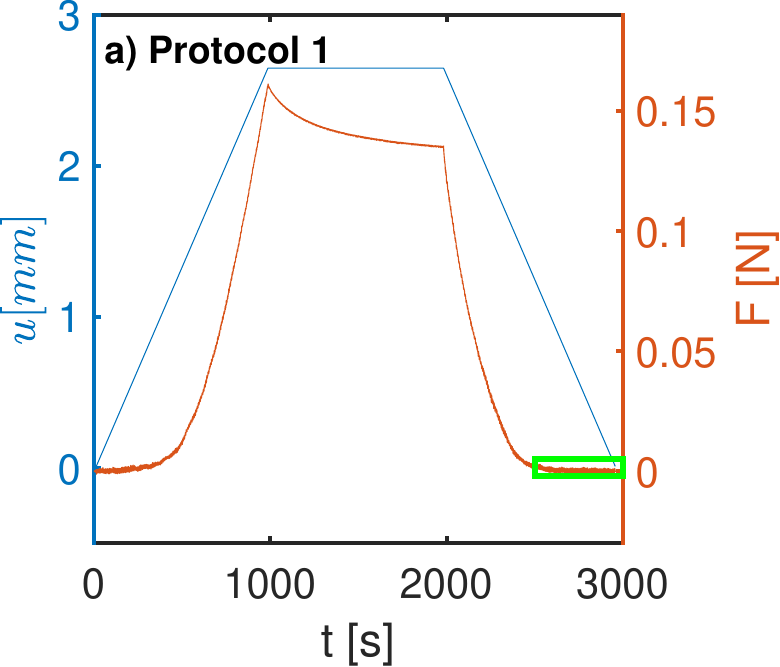}
    \includegraphics[scale=0.29]{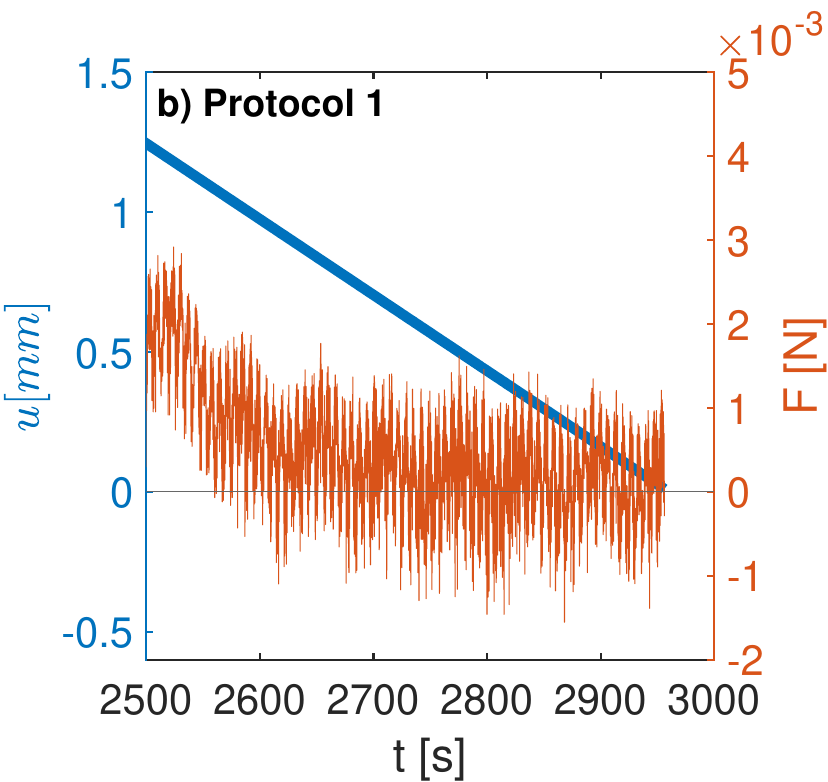}
    \includegraphics[scale=0.28]{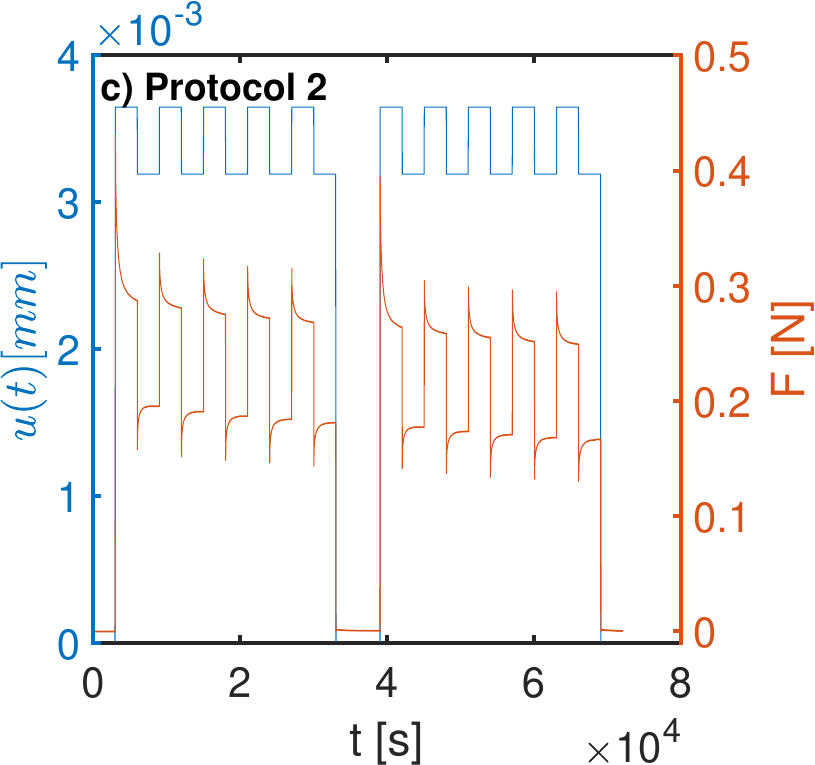}
    \includegraphics[scale=0.28]{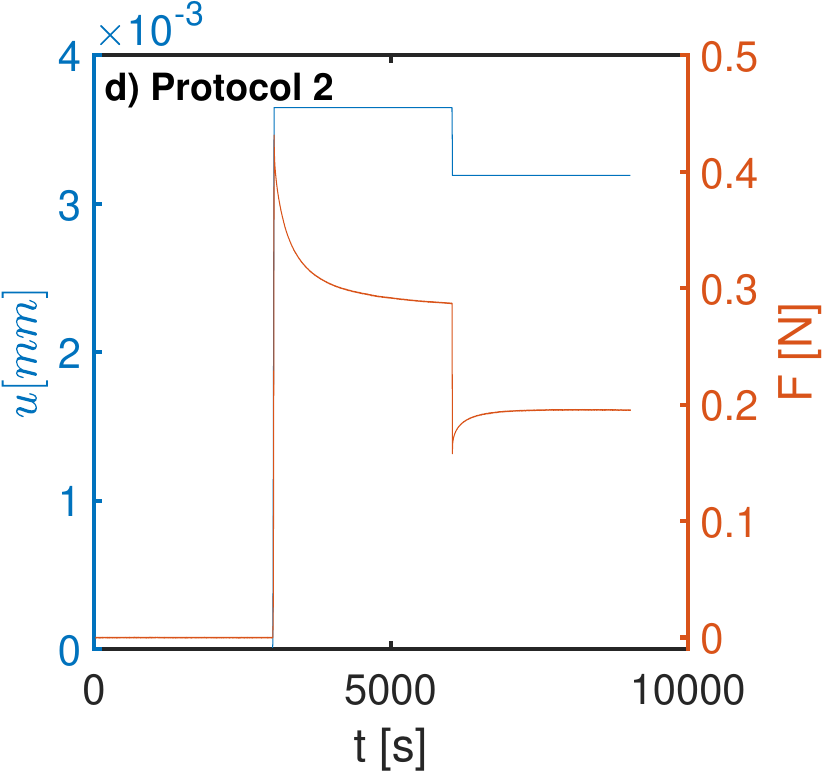}
    \includegraphics[scale=0.33]{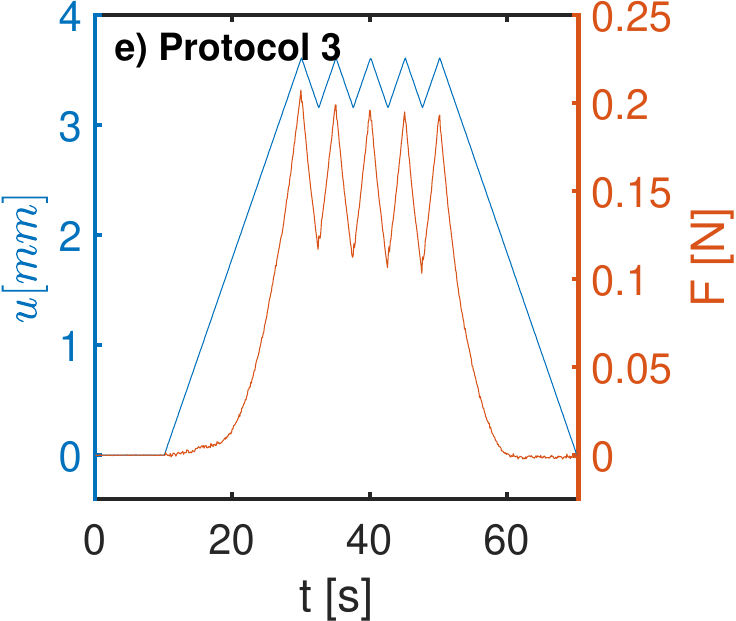}
    \includegraphics[scale=0.28]{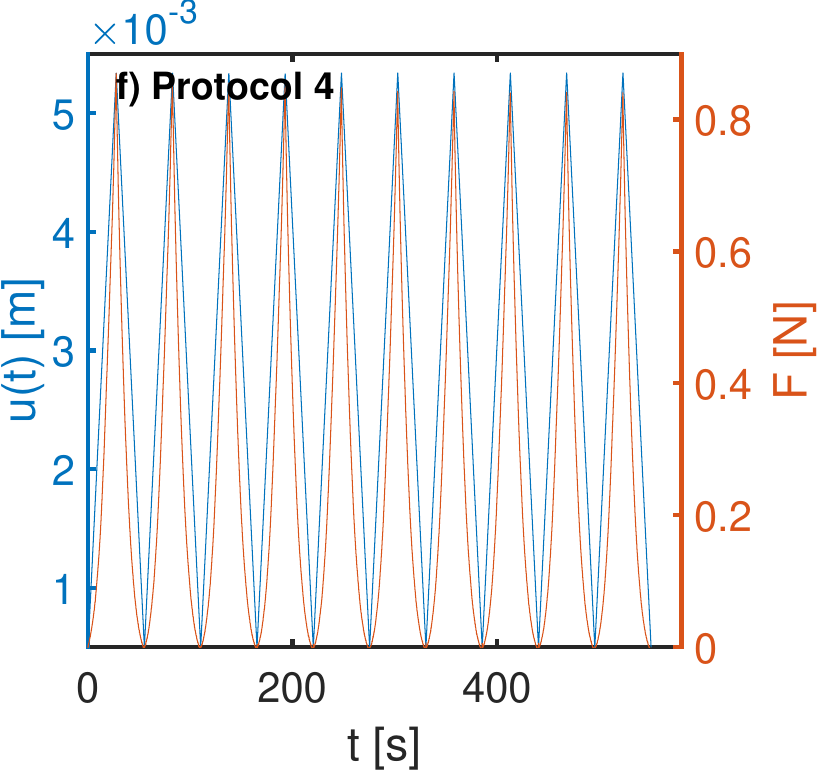}
    \includegraphics[scale=0.32]{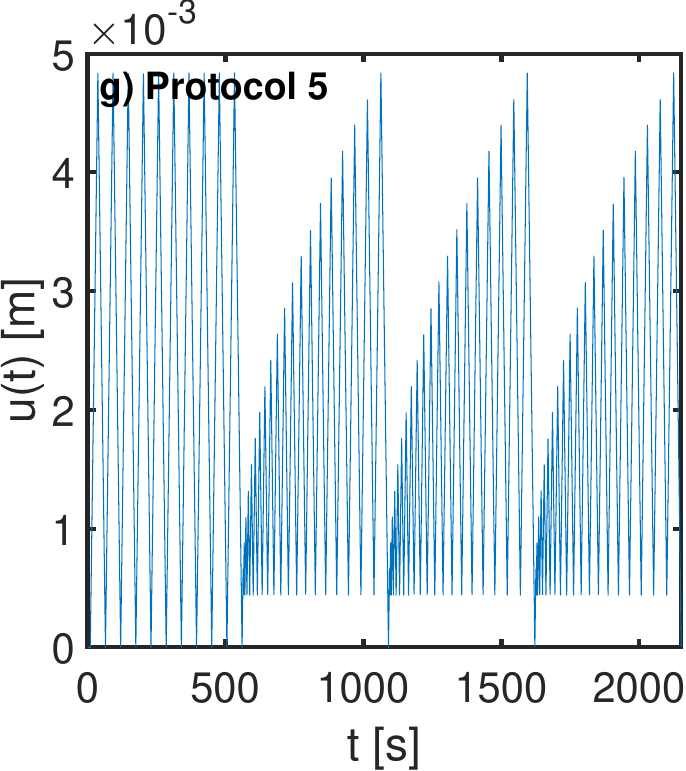}
    \includegraphics[scale=0.32]{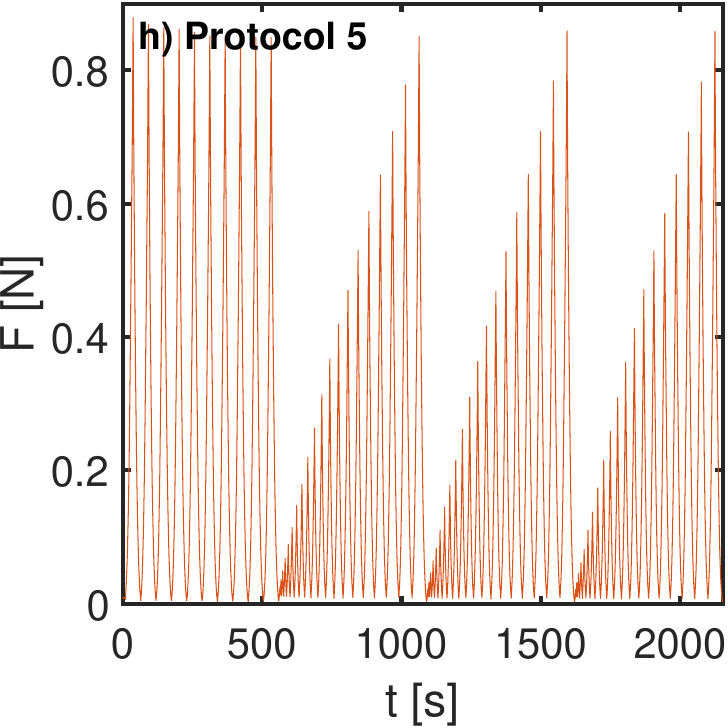}
    \includegraphics[scale=0.28]{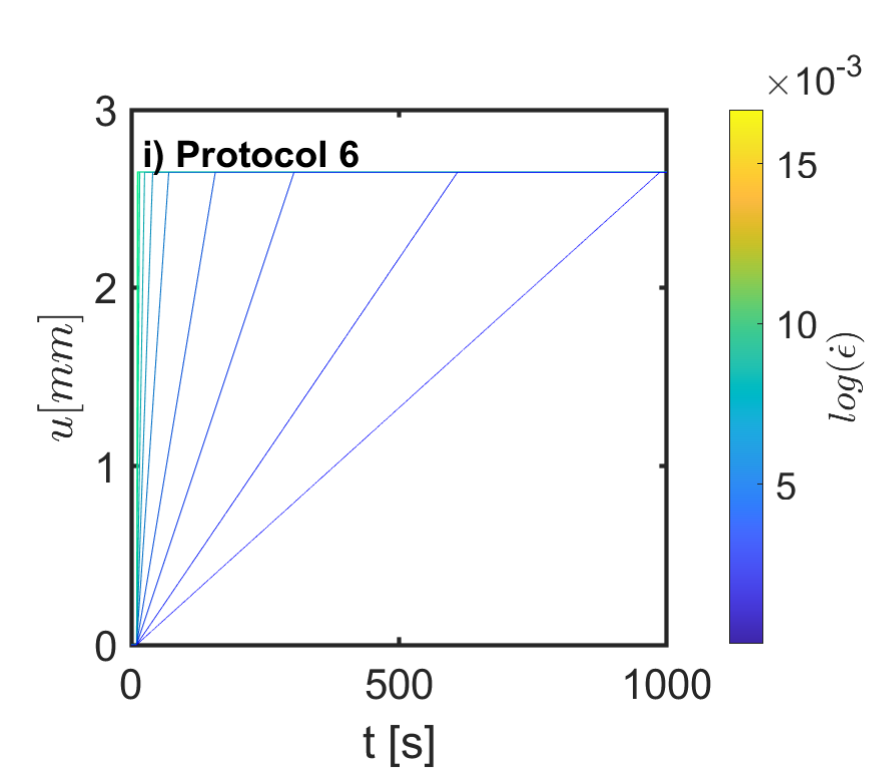}
    \includegraphics[scale=0.28]{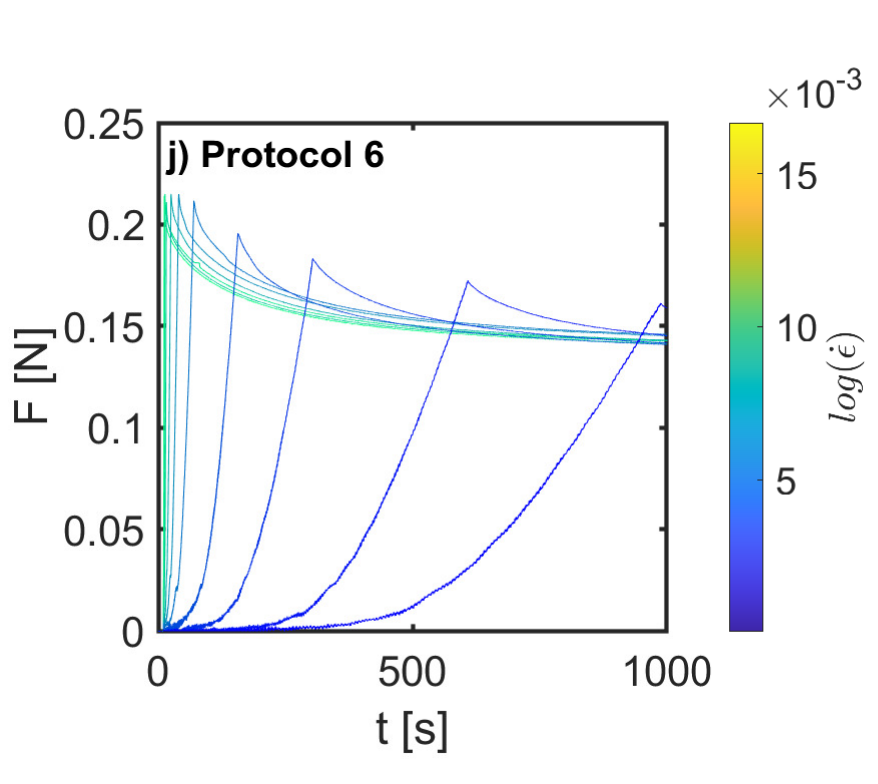}
    \caption{\label{fig:relaxation_diff_protocol}
    Experimental results for various protocols described in Sec.~\ref{exp_protocols} for a 3D packing of hydrogel spheres. a) Strain and force for Protocol 1. The imposed strain is in blue and the output force response is in orange. The green box is magnified in panel b): magnified results for strain and force as a function of time during the late stages of decompression to $\epsilon=0$ for the protocol 1. c) Strain and force for Protocol 2. Note the difference in time scale as compared to panel (a).  d) Magnified results for the first cycle for Protocol 2. e) Strain and force for Protocol 3, f) Strain and force for Protocol 4. g) Applied strain for Protocol 5. h) Force response for Protocol 5. i) Applied strain for Protocol 6. j) Force response for Protocol 6.}
\end{figure}

This protocol consists of application of a single, finite rate  strain controlled compression-decompression cycle, with a wait step included. The test is designed to test the relaxation timescale of the packing at a constant deformation rate. For this protocol the packing is compressed to a specific $\epsilon$ at a specific $\dot\epsilon$ and the decompressed at the same rate $\dot\epsilon$ to its original position with a waiting time in between ($t_{wait}$). Figure \ref{fig:relaxation_diff_protocol}a shows a version of this protocol where a packing of 88.15 mm is compressed to a $\epsilon$ of 3 percent strain (compression depth=2.645 mm) with a $t_{wait}$ of 1000 [sec] and $\dot\epsilon$ of $3\times 10^{-5}[sec^{-1}]$ and decompressed at the same $\dot\epsilon$ to a $\epsilon$ of 0 in blue and the measured force response ($F$ [N]) in orange. the packing height was 88.15 mm. Figure \ref{fig:relaxation_diff_protocol}b shows the magnified region of the tail end of the experiment shown in the green window shown in green. In this figure, we see that there are no negative or tensile forces at the end of the decompression cycle when $\epsilon=0$, except for signals that come from detector noise.  

\subsubsection*{\label{incomplete_compression_withrelaxation} Protocol 2: Relaxation repeatability test}
This protocol consists of a series of relaxation tests when the packing is strained and then unstrained repeatedly, at a finite offset strain. The protocol was designed to test the repeatability of the relaxation timescales. In this protocol, the packing is compressed to a specific $\epsilon$ at a specific $\dot\epsilon$ and then decompressed at the same rate $\dot\epsilon$ but to a different finite strain level, with a wait time in between ($t_{wait}$). Figure \ref{fig:relaxation_diff_protocol}c shows a version of this protocol where a packing of 91.2mm is compressed to a $\epsilon$ of 4 percent strain (compression depth=3.65 mm) with a $t_{wait}$ of 3000 [sec] and $\dot\epsilon$ of $2\times 10^{-3}[{\rm sec}^{-1}]$ and decompressed at the same $\dot\epsilon$ to a $\epsilon$ of 3.5 percent in blue and the measured force response ($F$ [N]) in orange. We then repeat the experiment 4 times with $\epsilon$ varying between 4 and 3.5 percent before returning to $\epsilon=0$. The process is then repeated after waiting for 8000 seconds. Figure \ref{fig:relaxation_diff_protocol}d shows the first cycle of this experiment.

\subsubsection*{\label{incomplete_compression_withoutrelaxation} Protocol 3: Cycle test at finite strain}
To study the relaxation displayed by the packing under continuous strain cycling at a finite overall strain value we develop Protocol 3. This test is similar to Protocol 2, but without the wait steps. For this protocol, the packing is compressed to a specific $\epsilon$ at a specific $\dot\epsilon$ and then decompressed at the same rate $\dot\epsilon$ but at a different strain level than 0 with no waiting time between. Figure \ref{fig:relaxation_diff_protocol}e shows a version of this protocol where a packing of 90.3mm is compressed to a $\epsilon$ of 4 percent strain (compression depth=3.61 mm)  at a $\dot\epsilon$ of $2\times 10^{-3}$ and decompressed at the same $\dot\epsilon$ to a $\epsilon$ of 3.5 percent in blue and the measured force response ($F$ [N]) in orange. This protocol does not have a waiting time in between the loading and unloading phases.

\subsubsection*{\label{cyclic_compression}Protocol 4:  Cycle test to zero strain}
This protocol was designed to study relaxation in the packing under continuous cyclic strain load as well. However, in this protocol the packing was unstrained fully so that the complete dynamics and hysteresis in the force response could be studied. For this protocol, the packing is compressed from $\epsilon=0$ to a specific $\epsilon$ at a specific $\dot\epsilon$ and then decompressed at the same rate $\dot\epsilon$ to $\epsilon=0$ with no waiting time between.  The cycle is repeated multiple times. Figure \ref{fig:relaxation_diff_protocol}f shows this protocol where a packing of 90.3mm is compressed to a $\epsilon$ of 5.5 percent strain (compression depth= mm)  at a $\dot\epsilon$ of $2\times 10^{-3}$ and decompressed at the same $\dot\epsilon$, with applied strain in blue and the measured force response ($F$ [N]) in orange.

\subsubsection*{\label{incremental_cyclic_compression} Protocol 5: Incremental cyclic amplitude test}
This protocol was designed to verify the strain dependence of force dynamics and force hysteresis. For this protocol, we first start with a preparatory step. The packing is compressed from $\epsilon=0$ to a specific $\epsilon$ at a specific $\dot\epsilon$ and then decompressed at the same rate $\dot\epsilon$ to $\epsilon=0$ with no waiting time in between. The cycle is repeated 10 times but the data is not used for interpretation. Then, the packing is compressed to a strain of $\epsilon=0.005$ which acts as a baseline $\epsilon_b$. The packing is then compressed from $\epsilon_b$ to a $\epsilon$ 0.01 and back to $\epsilon_b$. Then the packing is compressed to a higher $\epsilon$ 0.015 and back to $epsilon_b$. Strain is applied at increments of 0.005 and decompressed back to $\epsilon_b$ to a maximum strain of 0.055. The entire experimental stage is repeated 3 times. Figure \ref{fig:relaxation_diff_protocol}g shows the applied displacement($u$) from this protocol, where a packing of 90.3~mm is compressed to a $\epsilon_{max}$ of 5.5 percent strain (compression depth= mm)  at an $\dot\epsilon$ of $2\times 10^{-3}$ and decompressed at the same $\dot\epsilon$ to an $\epsilon_b$ of 0.005 percent. Figure \ref{fig:relaxation_diff_protocol}h shows the measured force response.

\subsubsection*{\label{variable_strain_rate}Protocol 6: Relaxation test for different strain rate}
This protocol was designed to test strain rate sensitivity of the relaxation test described in Protocol 1. The sample is compressed to an $\epsilon$ of 3 percent and unloaded to $\epsilon$ of 0 percent with 1000 seconds of waiting time in between the loading and unloading phases. However, for this protocol, the $\dot\epsilon$ is varied. The experiments were conducted from the slowest experiments done first. The strain rates tested ranged between $3\times10^{-5}$ and $1.67\times10^{-2}$ [/sec]. The applied strain $\epsilon$ vs time plots can be viewed in Fig.~\ref{fig:relaxation_diff_protocol}i. Figure \ref{fig:relaxation_diff_protocol}j the corresponding force responses to the applied strains. Here, the color bar represents the log of the applied strain rate to properly appreciate the range of strain rates used. The packing depth for this protocol was 88.15 mm.

\begin{figure}
    \centering
    \includegraphics[scale=0.4]{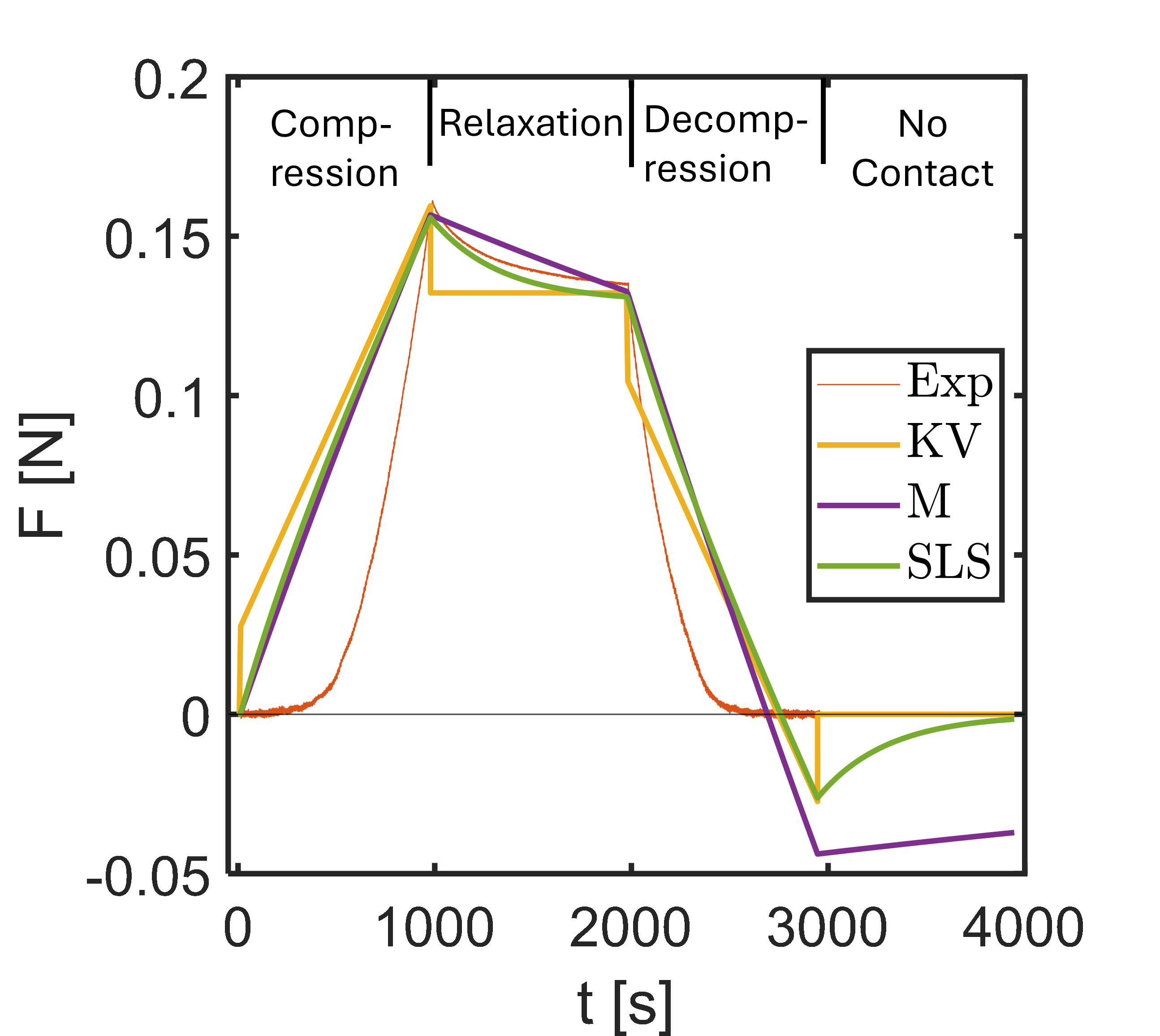}
    \caption{Diagrammatic representations of force response against time under applied displacement protocol 1 (Fig.~\ref{fig:relaxation_diff_protocol}b. Responses from some traditional models are overlaid against the response seen from the experiment (orange). This includes 1) The discontinuous response of the KV model (yellow). 2) The response of the Maxwell (M) model to a constant non-null displacement asymptotically tends to zero. (Purple) 3) The response of the Solid Linear Solid (SLS) model(Green) can be negative under null imposed displacement.}
    \label{fig:tradmodel}
\end{figure}

\section{\label{new_model}Strain Dependent ViscoElastic Solid Model (SDVES)}
There are many phenomenological models that can be used to capture the response of a material that is exposed to the experimental protocols described in Sec.~\ref{exp_protocols}. We have mentioned some of these models in Sec.~\ref{intro}. Some of the simplest and phenomenological models that account for time- and strain rate dependent material are viscoelastic models, represented by some combination of springs and dashpots. These include the usual Kelvin-Voigt (KV), Maxwell (M), and standard linear solid (SLS) model. To introduce some notation and as a brief reminder of the functionality of these models, we refer to Appendix~\ref{tradmodel}. However, based on the observations mentioned in the Fig.~\ref{fig:tradmodel}, we can immediately conclude that none of the simpler models capture all the major characteristics of the compression response of a cohesionless soft granular packing composed of hydrogels.

\subsection{Comparing standard viscoelastic models}
We show a representation of the Kelvin-Voigt, Maxwell, and the standard linear solid model in Fig.~\ref{fig:tradmodel},  overlaid against the experimental force response for the applied displacement protocol described in Protocol 1 (Sec.~\ref{complete_compression}). It is straightforward to identify the qualitative mismatches between models and experimental data. The KV model cannot replicate the non-linear increase in force in the compression stage, the non-linear decay of force during relaxation and decompression, and predicts a tensile force as we approach the loss of contact. The Maxwell model would allow the force to decay asymptotically to zero at a finite applied displacement. If the material were to be allowed to relax for a long enough time, during the relaxation phase, it would produce a negative force towards the end of the contact during decompression, and would predict a stretching of the contact. The SLS model only fails at the decompression phase, as it also predicts an adhesive force, which is unphysical as the particles are noncohesive when fully submersed.\\ 

\begin{figure}[!thb]
    \centering
    \includegraphics[scale=0.4]{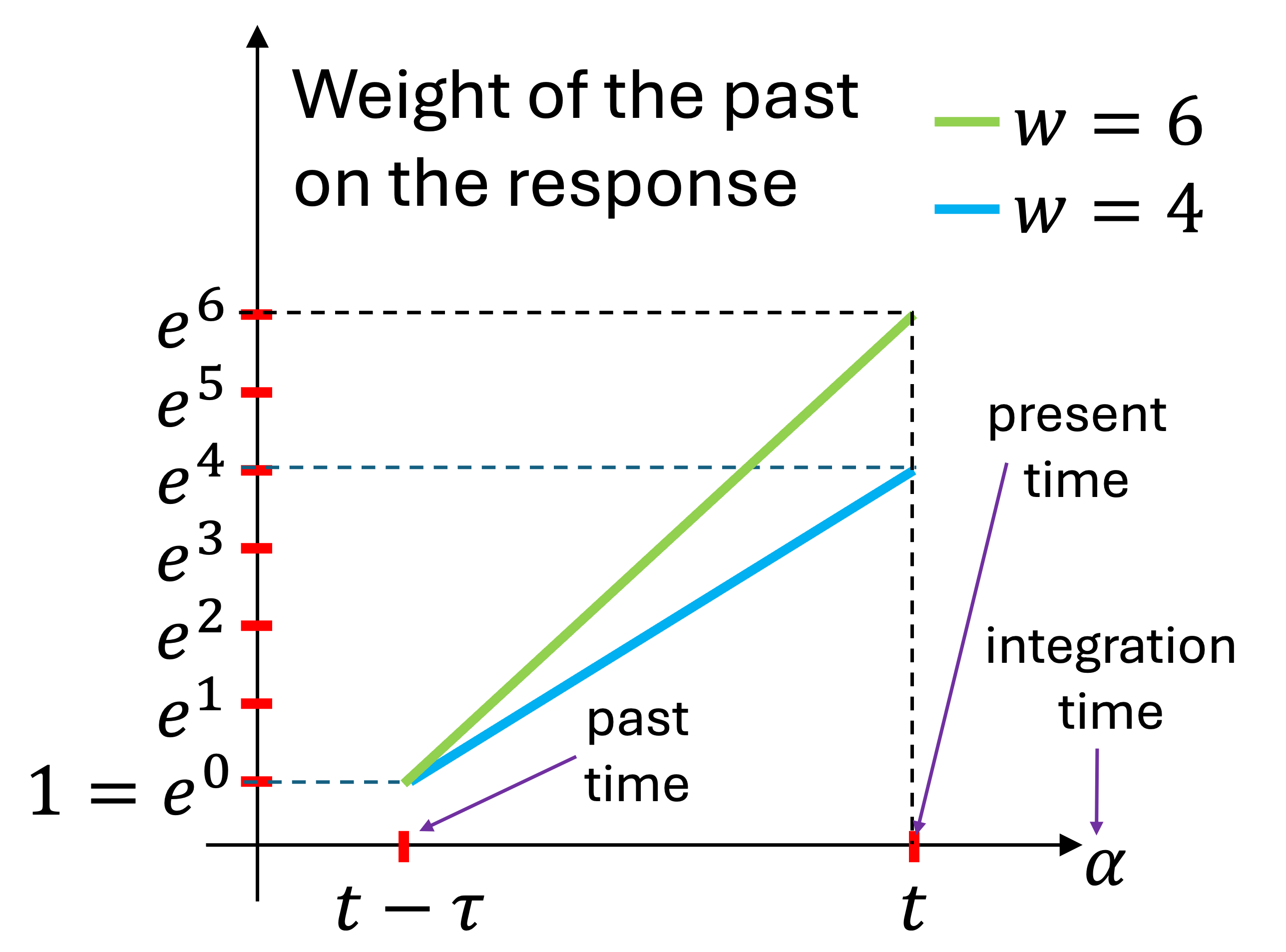}
\caption{ \label{weightofthepast} The weight of the past for two values of $w$, i.e. for $w=4$ and $w=6$ is shown in semilogarithmic scale, as incorporated in the SDVES.}
\end{figure}

\begin{figure*}[!thb]
    \centering
    \includegraphics[scale=0.4]{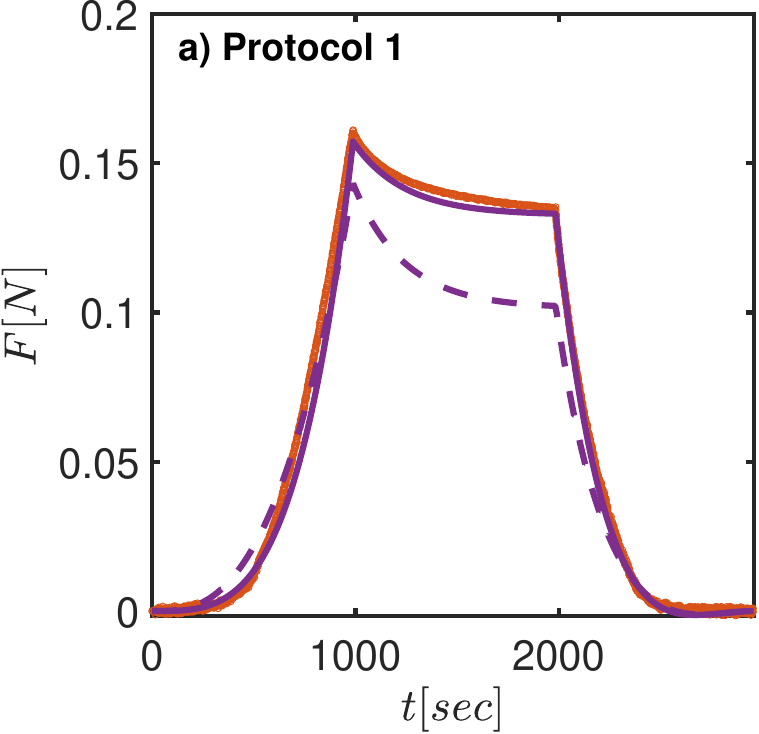}
    \includegraphics[scale=0.4]{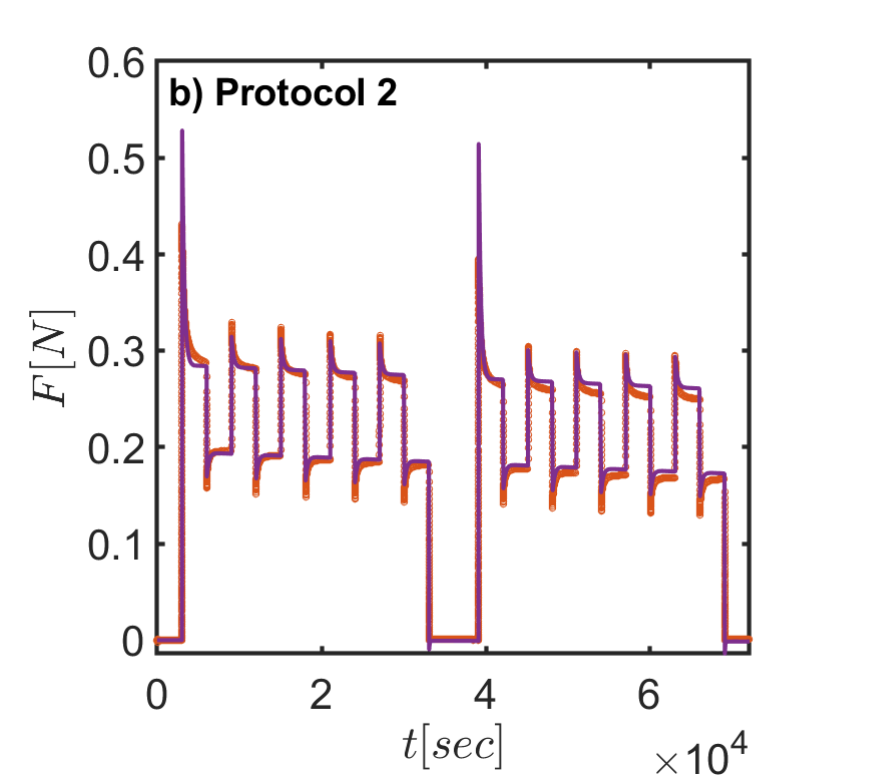}
    \includegraphics[scale=0.4]{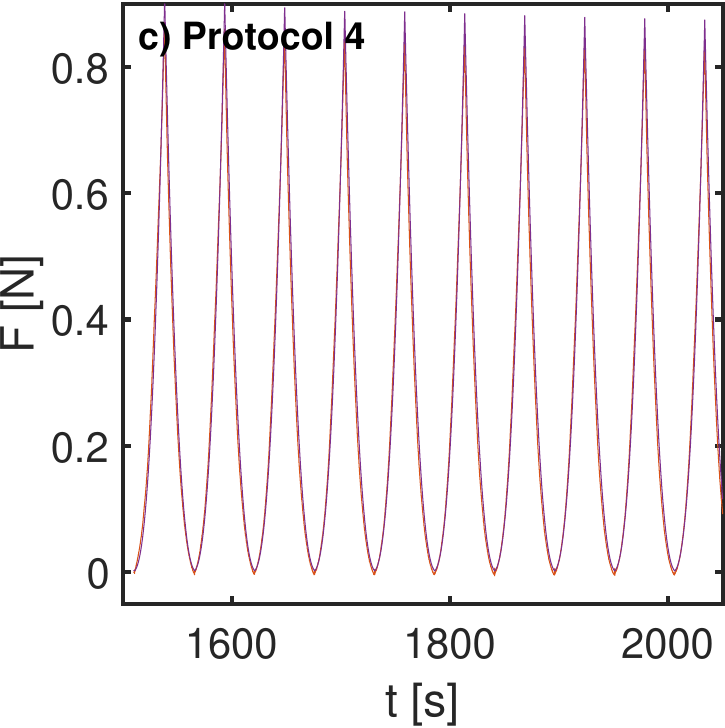}
    \includegraphics[scale=0.4]{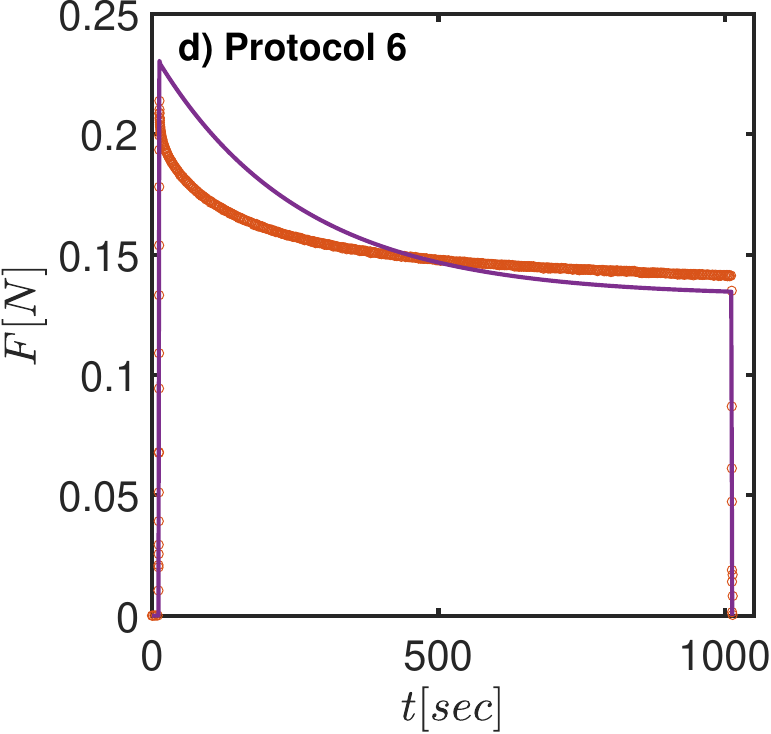}
    \includegraphics[scale=0.45]{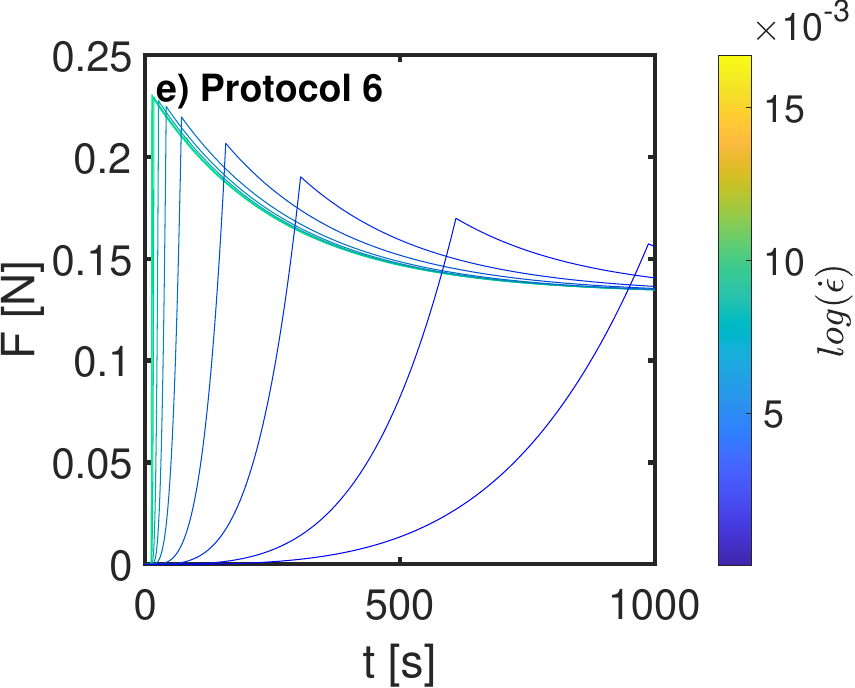}
    \includegraphics[scale=0.4]{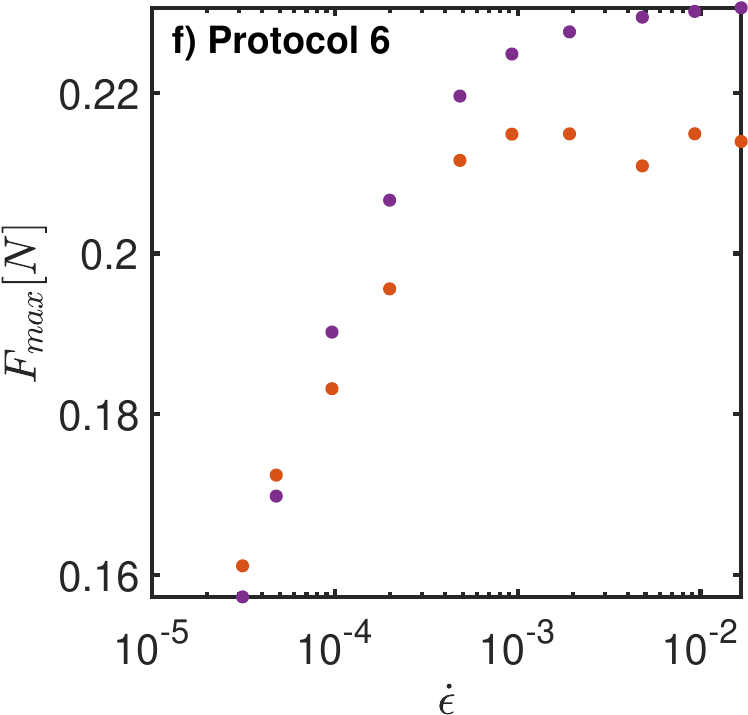}
    \includegraphics[scale=0.4]{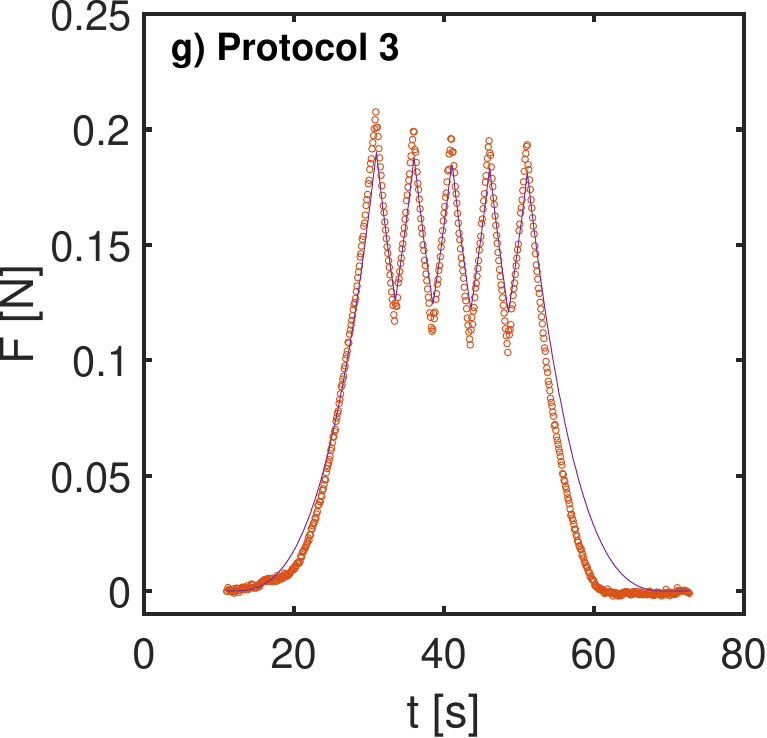}
    \includegraphics[scale=0.4]{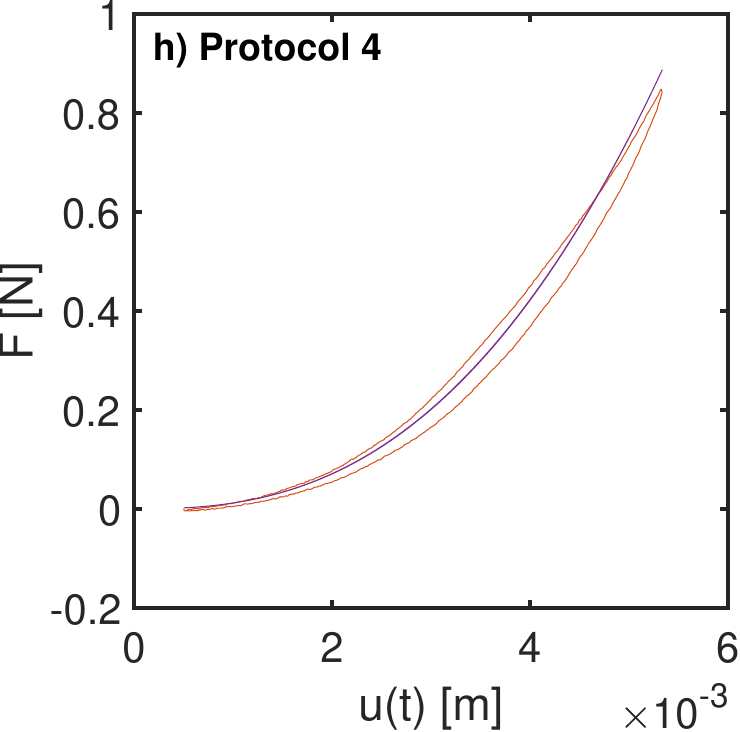}
    \includegraphics[scale=0.4]{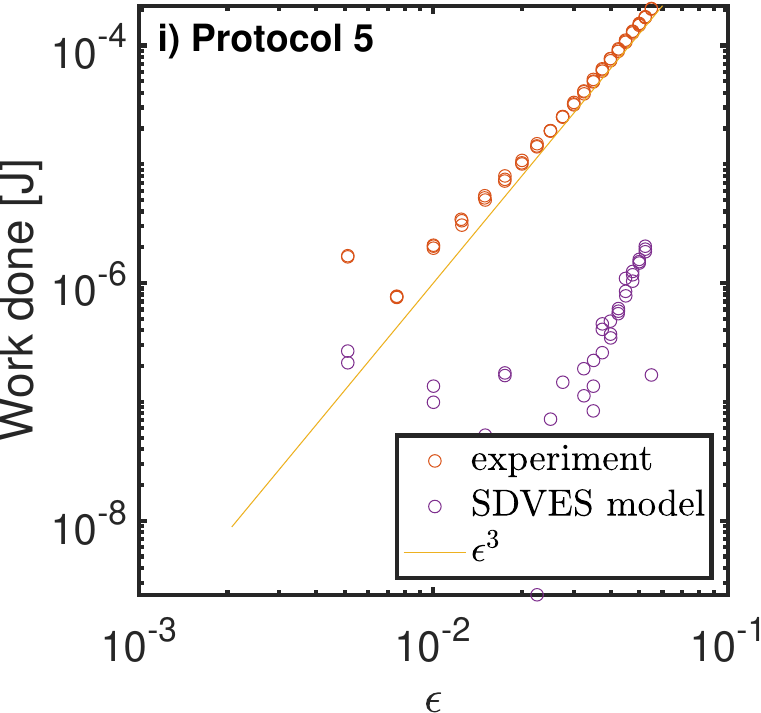}
\caption{ \label{good_fits} Experimental force data (orange circles) and SDVES fit results (lines)  for different experimental protocols described in Sec.~\ref{exp_protocols} for a 3D packing of hydrogel spheres. a) Protocol 1, with best fit (solid line),  with parameter set as used in panel b) (dotted line). b) Protocol 2 with best fit. c) Protocol 4 fitted with he parameter set as described in table \ref{parameter_table}. d) Protocol 5. e) Predicted SDVES model responses for the Protocol 5, to be compared with experimental results shown in Fig.~\ref{fig:relaxation_diff_protocol}j. f) Maximum force from the experiment and model predictions vs. the applied strain rate $\dot\epsilon$ for Protocol 5. g) Results for Protocol 3.  h) Protocol 4. i) the work done after loading and unloading vs. maximum strains applied using Protocol 5. The model parameters for all fits used can be found in Table \ref{parameter_table} in the Appendix.}
\end{figure*}

\subsection{Introducing a non-linear strain dependent correction to SLS}
To overcome the shortcomings of the standard visco-elastic models, we introduce a modified strain dependent viscoelastic solid model (SDVES). Here, the combination of an elastic and viscous response is the parallel of a nonlinear spring and a strain dependent Maxwell element. The relative displacement $u_N$ is therefore common to that of the nonlinear spring and to the strain dependent Maxwell element.
The reaction force $F_{SDVES}$ of the new model is the sum of that of the nonlinear spring $F_{Ns}$ and of the strain dependent Maxwell element $F_{NM}$,
\begin{equation}\label{N1}
    F_{SDVES}=F_{Ns}+ F_{NM}.
\end{equation}
The reaction of the nonlinear spring is nonlinear with respect to $u_N$,
\begin{equation}\label{N2}
    F_{Ns}=k_N (u_N)^{a}
\end{equation}
Here, $a$ is an exponent chosen to represent the nonlinear response in $F_{Ns}$ with $u_N$. $k_N$ is an effective proportionality constant that represents the stiffness of the elastic response. Since $F_{Ns}$ needs to be a unit of Force [N], and $u_N$ takes the unit of distance [m], the unit of the constant $k_N$ will depend on $a$.
Furthermore, the reaction of the strain dependent Maxwell element is inspired by the integral solutions (Eq.~\ref{m5}) of the Maxwell and of the standard linear solid model (Eq.~\ref{st7}),
\begin{eqnarray}\label{N3}
    F_{NM}(t)&=&2\frac{c_N}{\tau^2}\left(\frac{u_N (t)}{u_{c}}\right)^b \int_{t-\tau}^t \Biggl\{ \Bigl(u_N(t)-u_N(\alpha) \Bigl) \\ \times  
    && \exp\left( w \frac{\alpha-t+\tau}{\tau} \right) \Biggl\} d \alpha. \nonumber
\end{eqnarray}
Here, $\tau$ is a relaxation time that represents the timescale of the exponential force decay in the viscous part of the model. This parameter controls how long the effects of a given load are expected to be observed.
It is similar to the relaxation time $\tau_M=c_M/k_M$ (or $c_{stM}/k_{stM}$) in Eq.~\ref{m5} or Eq.~\ref{st7} of the Maxwell (or Standard) model response.
In our model, the time scale also controls how long of a load history will affect the mechanical response of the material at any given time because the duration of the integration in time in Eq.~\ref{N3} is given by the constitutive quantity $\tau$.
This solves the paradoxical situation, present in Maxwell and Standard models, in which the influence of the history of the displacement load stands for a time duration that is very small at the beginning of the response and very large at its end.\\ 

\subsection{The nature of the history dependence}
Let us now discuss the meaning of the parameter $w$ with the help of Fig.~\ref{weightofthepast}.
The exponential form of the kernels of the integral response of the Maxwell (Eq.~\ref{m5}), the Standard Linear solid (Eq.~\ref{st7}) and of the proposed (Eq.~\ref{N3}) models are clear from their associated equations. Each of these exponents carry a kernel in the integral form.
These kernels weigh the contribution of the displacement history evaluated at $\alpha$ within the time integration range on the reaction force response at time $t$.
In particular, these kernels say that the displacement at the present time at $\alpha=t$ weighs more than in the past, which is represented by $\alpha=0$ for Maxwell and Standard models and by $\alpha=t-\tau$ for the proposed model.
However, the ratio of these weights in both the Maxwell and the Standard models is the Napier's constant $e=\exp(1)$ and it is the same for all materials.
We avoid this arbitrary prescription in the new model, and the weight factor is constitutively prescribed to be $\exp(w)$, see Fig. \ref{weightofthepast}.
$c_N$ is a prefactor that determines the strength of the viscous response (Eq.~\ref{N3}) compared to the elastic response (Eq.~\ref{N2}). We would like to highlight that Eq. ~\ref{N3} can be rewritten in a different way that highlights how such description can be used to change the integration of the history dependence of the model: 

\begin{eqnarray}\label{N3-rewrite}
    F_{NM}(t)&=&2\frac{c_N}{\tau^2}\left(\frac{u_N (t)}{u'_{c}}\right)^b \int_{t-\lambda}^t \Biggl\{ \Bigl(u_N(t)-u_N(\alpha) \Bigl) \\ \times  
    && \exp\left(\frac{\alpha-t}{\tau} \right) \Biggl\} d \alpha. \nonumber
\end{eqnarray}
Here, the $\exp(b)$ has been absorbed in the new material constant $u'_c$ and a cutoff time $\lambda$ has been introduced to reflect the existence of a history cutoff timescale; either way, a second time scale needs to be introduced to capture the observed relevance of the history of the material. 

The term $(u_N(t)/u_c)^b$ in Eq.~\ref{N3} is responsible for the ``strain-dependent'' aspect in SDVES. If the displacement $u_N$ is equal to the characteristic displacement $u_c$, then the term has no role. However, if the displacement $u_N<<u_c$ is smaller than the characteristic one, then the viscous contribution to the response becomes negligible and the higher the exponent $b$ is the lower is the contribution of the viscous term on the response. As a consequence, this combination of factors ensures that we can choose paramaters that ensure that forces are not attractive at the end of the decompression, consistent with our experimental data, and a feature that cannot be replicated from any of the other traditional models. 
Another important characteristic of the non-local (in time) constitutive relation (Eq.~{N3}) is that for a very small characteristic time $\tau$, i.e. in the limit of the characteristic time $\tau$ to zero, Eq.~\ref{N3} tends to
\begin{equation}\label{N4}
    \lim_{\tau \rightarrow 0} F_{NM}(t) = c_N \dot{u}_N \left( \frac{u_N}{u_{c}}\right)^b.
\end{equation}
Thus, for $b=0$ and $a=1$, the insertion of Eqs.\ref{N2} and \ref{N4} into Eq.~\ref{N1} yields a reaction force that is the same as that from the Kelvin-Voigt model illustrated in Eq.~{kvmodel} with stiffness $k_N$ and viscosity $c_N$.
Besides, $u_{c}$ is the characteristic displacement scale chosen so that the ratio before the integral sign in Eq.~\ref{N3} or in Eq.~\ref{N4} is always lower than unity. 
Such a ratio makes the dissipating contribution $F_{NM}$ to the reaction force $F_N$ negligible for small values $u_N << u_{c}$ of the imposed displacement $u_N$.
The reason for such a factor is to avoid the negative force that is experienced in the Maxwell model for an imposed displacement that is initially positive and asymptotically null.

\section{\label{interpretation}Comparing SDVES with experiments}
We can now compare the results of the SDVES model with the various experimental protocols mentioned in the Sec.~\ref{exp_protocols}. We start from Protocol 1. The SDSL model performs satisfactorily, as shown in Figure \ref{good_fits}a, where we plot the experimental force response. As can be seen from the solid purple line in Fig.~\ref{good_fits}a, the new model can replicate the nonlinear force growth during the compression stage, the force decay during the static applied strain stage and the absence of tensile forces during the decompression stage with a single set of parameters in all three phases for the protocol. A comparison with Fig.~\ref{fig:tradmodel} shows that SDVES performs much better than simpler models.\\ 
Figure \ref{good_fits}b shows the experimental data and the model fits from Protocol 2. As can be seen from the figure, the model is capable of predicting the force response over compression, static strain and decompression over multiple cycles, although small deviations can be observed in the relaxation step, which can be identified with the simplicity of a single relaxation timescale approach used here. The model even predicts the recovery of the peak force after waiting for $t>\tau$ at $\epsilon=0$. Figure \ref{good_fits}c shows the comparison of the experimental data and the model response for the cyclic protocol described in Sec.~\ref{cyclic_compression}. Here, we see that the experimental results and the model predictions are comparable.
We note that the parameters used in the simulation of the three protocols vary. The parameters used for figure \ref{good_fits}a is different from those used in \ref{good_fits}b and c, even though the material used for the different protocols should be the same. The dotted line in figure \ref{good_fits}a plots the model's prediction to protocol 1 with the model parameters used in figure \ref{good_fits}b. This variability of model parameters shows that there are some aspects of the SDVES model that give room for improvement. 

First, Figure \ref{good_fits}d shows the force response from the experiment and the predicted forces described Sec.~\ref{variable_strain_rate} for a different strain rate $\dot\epsilon=2\times10^{-3}$ while using the same fit parameters as shown for the solid line in figure \ref{good_fits}a. Here we see that the experimental results and the model predictions do not align in the relaxation behavior. However, if the parameters are modified, better fits can be achieved: the SDVES model works, yet the fit parameters, which should represent material constants, are not strain rate independent. The strain rate dependence is further investigated in \ref{good_fits}e,f where we compare the experimental and force responses via Protocol 6. Figure \ref{good_fits}e plots the predicted model force responses for various strain rates that can be compared to figure \ref{fig:relaxation_diff_protocol}i. Here, we see that the SDVES model matches the experimental responses at low applied strain rates, but starts to deviate slightly at higher strain rates. Figure \ref{good_fits}f plots the maximum forces at different $\dot\epsilon$ and the same maximum $\epsilon$ described in Sec.~\ref{variable_strain_rate}. Here we see that the peak forces after compression flatten out at higher strain rates in the experiment (red) but continue to grow in the model (purple). This is most likely due to the lack of a (nonlinear) strain rate dependent term in the model, making the force predictions of packing response deviate from experimental results at higher applied strain rates. We also note that the current model does not include the multitude of relaxation timescales observed in these packings~\cite{timescale-inprep}, providing further room for obvious model improvement.

Second, the predictions of the SDVES model underestimates the dissipation in the granular system at all applied maximum strains. Figure \ref{good_fits}g shows the force responses from Protocol 5, described in Sec.~\ref{incomplete_compression_withoutrelaxation}. The model captures both the initial compression and oscillatory compression quite well slightly but overestimates the force response during the decompression phase. This feature persists for the complete cyclic compression protocol described in Sec.~\ref{cyclic_compression}. The problem can be seen in figure \ref{good_fits}h when the compression curve is matched much better than the decompression curve. As a result, we see that the model underestimates the work done to complete a full cycle of compression and decompression in the packing at a $\dot\epsilon=2\times10^{-3}$ to a strain of $\epsilon=0.055$. We also observe that the energy dissipated in a single compression loop scales with a power law of approximately $\epsilon^3$ in the experiment. The overestimation of the force during decompression implies that the amplitude of this power law is underestimated. While de SDVES model does produce a power law scaling, the exponent seems to be inconsistent with the observations, as shown in \ref{good_fits}i which plots the energy dissipated as a function of maximum strain applied in the protocol described in Sec.~\ref{incremental_cyclic_compression}.

\section{\label{consistency} Stress control: 2D hydrogel disk packings}
To further test the SDVES model, we used an entirely different, stress-driven, experimental system to compare the model to. This quasi-2D system consists of $\sim$ 2000 bidisperse cylinders (with radii 120$\mu m$ and 160 $\mu m$ and height 90 $\mu m$, contributing equally to the volume faction) with parallel axes, confined in a microfluidic channel. Previously~\cite{timescale-inprep}, we found that, despite the differences in hydrogel material, particle size, dimensionality of the system, the system showed remarkably similar responses upon application of stress or strain, making this an ideal system to test the effectiveness of the SDVES approach.

\begin{figure}[!tbh]
    \centering
    \includegraphics[scale=0.2]{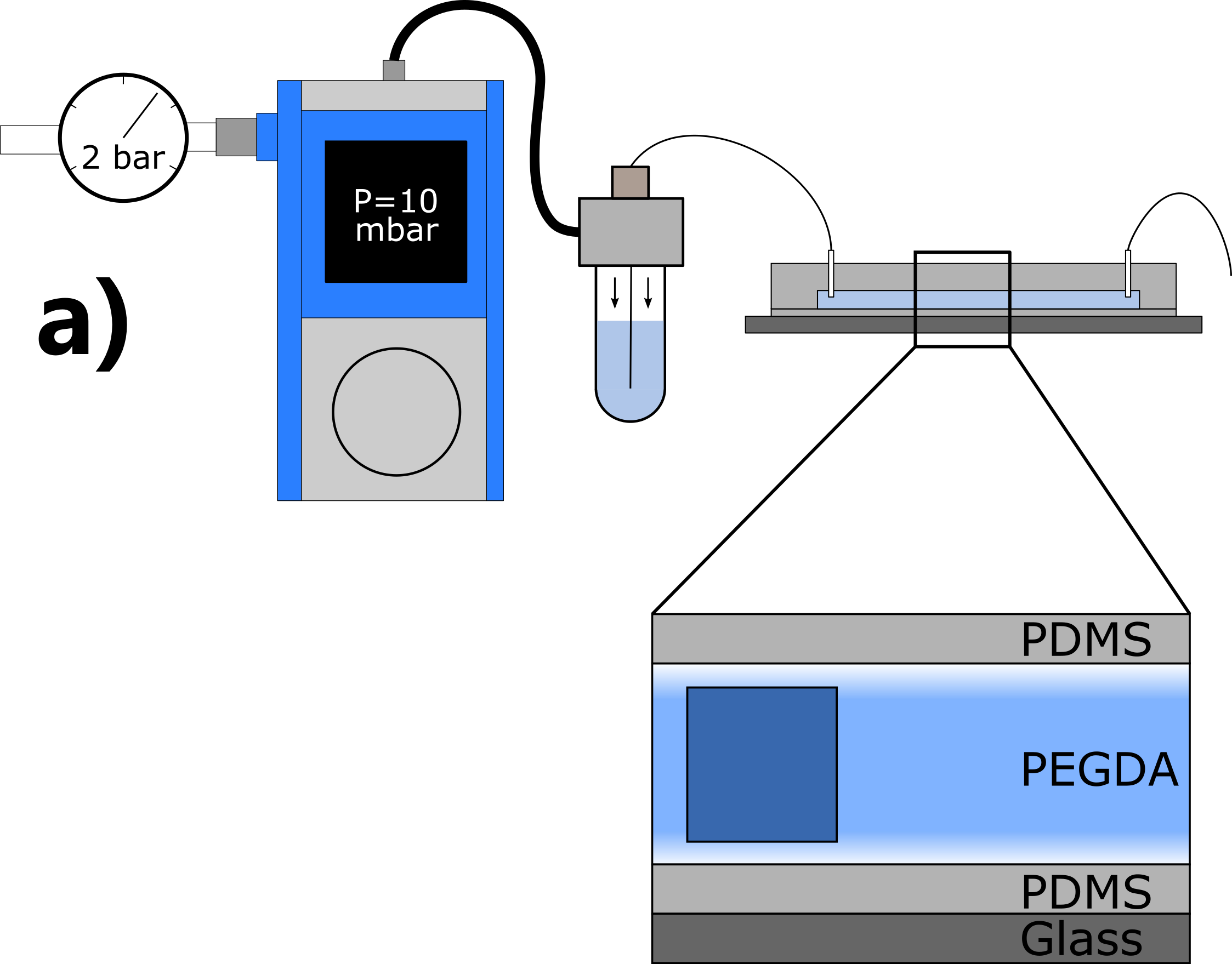}
    \includegraphics[scale=.7]{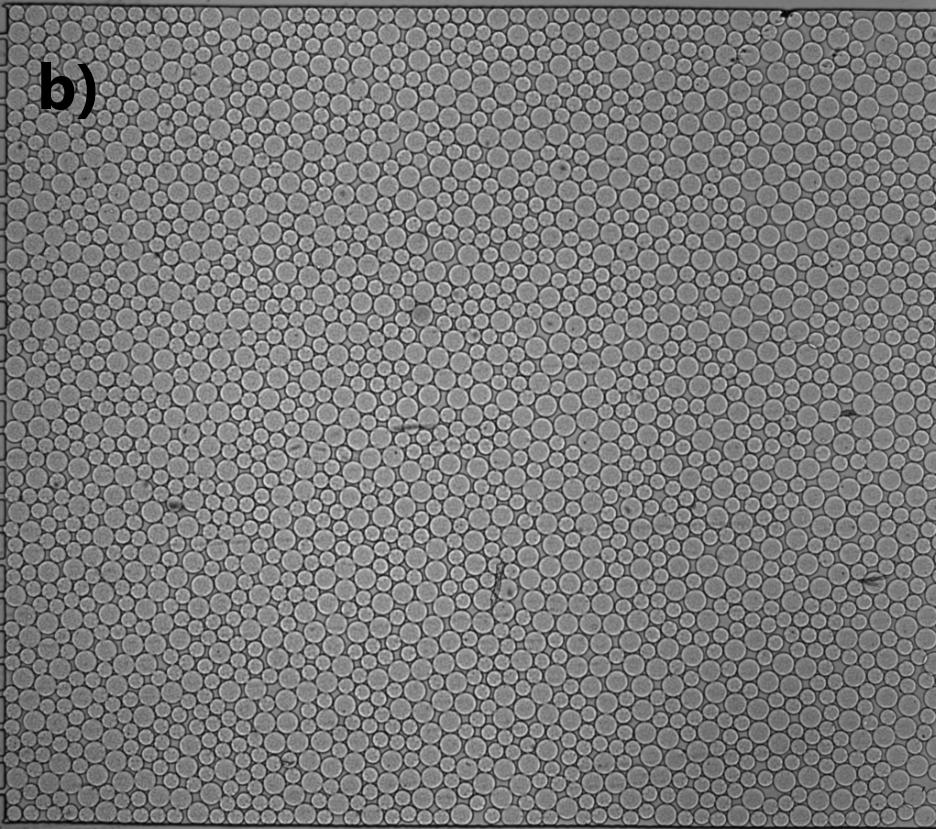}
    \includegraphics[scale=0.29]{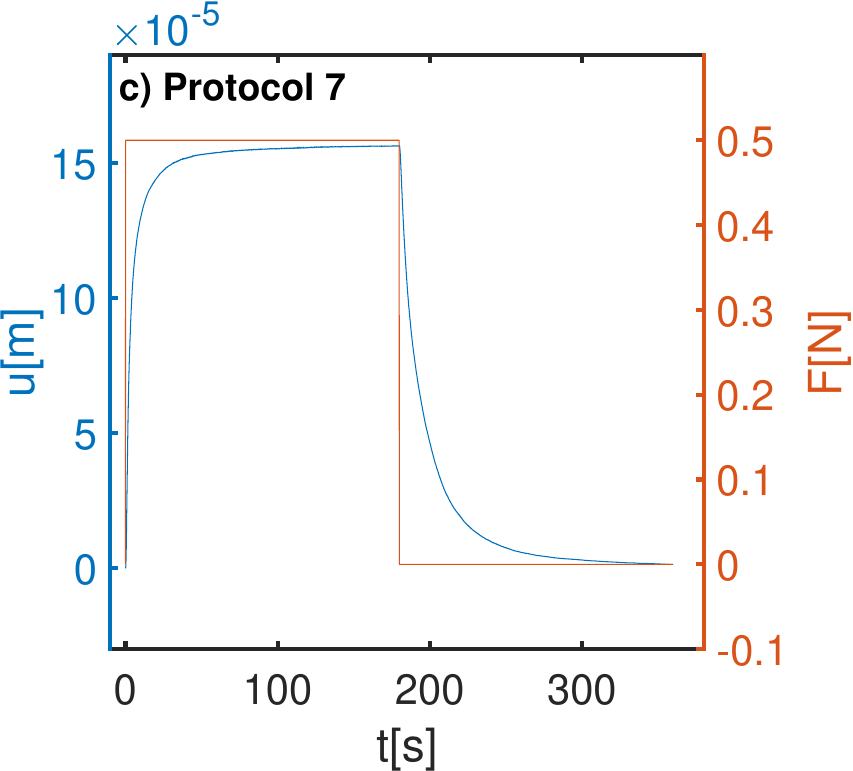}
    \includegraphics[scale=0.29]{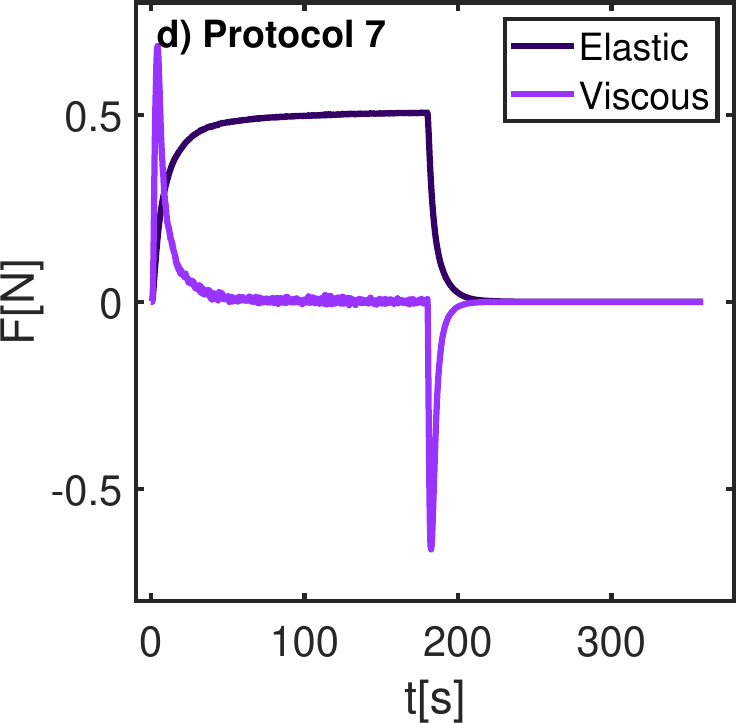}
    \includegraphics[scale=0.29]{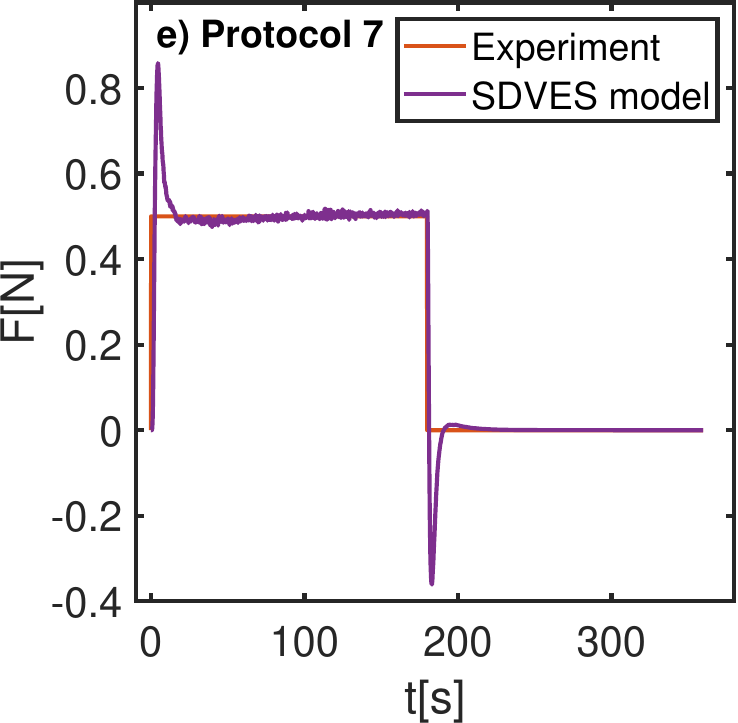}
    \includegraphics[scale=0.29]{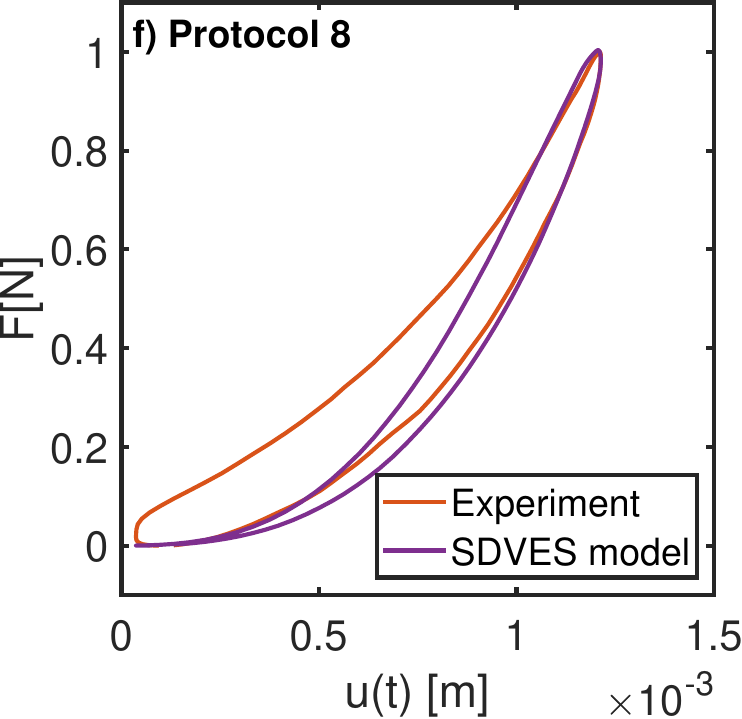}
    \caption{a) Schematic overview of the microfluidic quasi-2D compression experiment. b) Typically image obtained during an experiment. c) Strain response (blue) after a sudden increase and decrease in stress (red) at t=0 s and t=180 s, respectively. d) Elastic and viscous forces as calculated from the model, using the measured strain as input. e) Imposed force response (blue) and sum of elastic and viscous force as calculated by our model (orange). f) Force as function of strain during oscillatory compression, for a stress-driven experiment (blue), and our model (orange).}
    \label{fig:Stress_driven}
\end{figure}

\subsection{Sample preparation}
The methods largely follow the approach used described in~\cite{Kool2025}. Cylinders are fabricated in-situ by photo-polymerization of polyethylene glycol diacryclate (PEGDA) and a photo-initializer (2-Hydroxy-2-methylpropiophenone) \cite{dendukuri2006continuous, cappello2019transport}. The mixture in the microfluidic channel is locally exposed to UV-light (Osram HBO 103W/2 lamp with Chroma D365/10 filter) by blocking most of the light with a mask, where the opening in the mask determines the shape and size of the particle. The particles are made one-by-one, where a motorized XY-translation stage (Märzhäuser Wetzlar Tango PCI-E) moves the focal spot around the channel, indicated by a predetermined list of coordinates, obtained by simulating a bidisperse mix of particles settling in a box using LAMMPS.

The method creates cylinders slightly smaller than the channel height \cite{dendukuri2006continuous, cappello2019transport}, preventing the packing from buckling out of plane, hence, this packing can be considered quasi-2D. Similarly to the spherical poly-acrylamide hydrogels, the PEGDA cylinders are visco-elastic. However, their relaxation time is much faster, probably due to their poro-elastic features and smaller size~\cite{Biot1941GeneralConsolidation, Lin2006LoadSpace}.

The cylinders are fabricated in a channel containing a membrane which has pores of 50 $\mu m$, allowing the fluid to pass, while retaining the particles. Upon application of a hydraulic pressure via a Fluigent LineUp pressure controller, the fluid exerts a pressure on the particles, compressing them against the membrane, akin to a hydraulically driven piston. Note that pressure can be applied and removed quickly (the typical response time is 100-200 ms, to reach the newly set pressure). For a discussion on the viscous and normal stresses exerted by the fluid, please see Ref.~\cite{Kool2025}. What is relevant here is that the fluid flow around the disks allows us to very quickly modulate stresses exerted on the particle phase, while monitoring the strain per particle.

After particle fabrication, we sinusoidally compressed the particle packing against the membrane 10 to 20 times with a frequency of 1 cycle per 5 minutes and an amplitude of 10 kPa, to make sure the system has reached its densest configuration, eliminating any observed plastic strain due to compaction. After the cyclic compressions, the pressure was removed, and the packing was left to relax for 5 to 10 minutes to make sure each experiment was started in an undeformed state.

The deformation of the packing was determined by tracking each particle over time. This allows us to define the global strain as
\begin{equation}
    \varepsilon_{g}(t) = \frac{1}{N_{p}}\sum_{i}^{N_{p}}-\ln\left(\frac{x_{i}(t)}{x_{i}(0)}\right)
\end{equation}
where $N_p$ is the number of particles and $x_i(t)$ the position of particle $i$ along the compressive direction.

\subsection{Stress controlled experimental protocols}
\subsubsection*{Protocl 7: Step-stress} 
Similar to the creep tests for the 3D system (see Protocol 1 in Sec.~\ref{complete_compression}), we performed a step-like protocol, where at time t=0 the packing is compressed to a specific pressure $P$, and we followed the evolution of the deformation over time ($t_{wait}$ = 3 min). After which, we set $P=0$, and followed how the material recovers ($t_{wait}$ = 3 min). Again we note that in the 2D compression, we can switch $P$ fast ($<200 ms$, irrespective of $P$), unlike the 3D case where we are $\dot{\epsilon}$-limited. Images were collected at 10 Hz, to be able to capture the fast deformation at the start of the (de)compression. The tests in a pressure driven system can thus achieve high deformation rates, strongly testing the rate independent assumption built into the current SDVES model.

\subsubsection*{Protocol 8: Cyclic compression} 
The second protocol is a sinusoidal compression, following
\begin{equation}
    P(t) = -P_{a}\cos(2\pi f t) + P_{a}
    \label{eq:Oscillatory_compression}
\end{equation}
where $P_{a}$ is the pressure amplitude, and $f$ the oscillation frequency in Hz. Per oscillation cycle, 120 images were collected (irrespective of oscillation frequency) to limit the amount of data generated, while still able to track the particles. We found that the packing was in a transient state for $\sim 0.5$ cycle before reaching a limit cycle. The first cycle was used to establish the kernel history embedded in Eq.~\ref{N3} and required to compute predictions for the next cycle, but was not included in any further analysis. Protocol 8 is somewhat analogous to Protocol 4 used for the 3D hydrogel sphere packings.

\subsection{Results}
Inverting the SDVES model to take imposed force as an input and predict the observed deformation, poses a significant mathematical challenge, which is outside the scope of this research. However, we can easily check for consistency of the SDVES model with the stress-controlled data by using the observed strain as input for our model, and comparing the force output to the applied force. Figure \ref{fig:Stress_driven}c shows a typical strain response over time (blue) of the 2D packing upon a step-wise increase, followed by a step-wise decrease, in stress (orange). The measured strain response over time is then used to calculate the elastic (Figure \ref{fig:Stress_driven}d, blue) and viscous (Figure \ref{fig:Stress_driven}d, orange) components of the force, as per our model. 

The total predicted force, the sum of the elastic and viscous components, is shown in Figure \ref{fig:Stress_driven}e (orange), and compared to the actually imposed force (blue). We found that the model is able to predict the applied force reasonably well. The model, however, predicts an over/undershoot at the start of the compression/decompression, were to be expected, since a memory kernel introduces a characteristic response time. 

We also tested our model on a dynamic experiment, oscillatory compression around a mean pressure. Figure \ref{fig:Stress_driven}f shows the force-strain curve (blue) of the oscillatory compression, which bears resemblance to the Lissajous figures obtained from oscillatory shear rheology of viscoelastic materials like gels, however, with a non-linear elastic response. The force-strain curve predicted by our model is given in orange. We found that the model can reasonably predict the pressure (and therefore the dissipation) during a slow oscillation ($f = 1/600 \ Hz$) at large strains. At small strains, the model underestimates the contribution of the Maxwell element, resulting in an underestimated force and lack of dissipation. The presented SDVES model can clearly describe the observations of the stress driven compression experiments reasonably well, yet there are areas for future work. The stress control induces very high deformation rates, for which we had already observed the SDVES model to be less effective, as discussed in Sec.~\ref{interpretation}.



\section{\label{conclusion} Conclusions}
We perform strain- and stress-driven cyclic deformation tests on cohesionless soft granular packings made from hydrogel spheres or disks. We record both stress and strains and compare these mechanical observations with modeling efforts. Verifying that a range of standard (visco)elastic models is not capable of capturing the observations, we introduce a modified non-linear strain-dependent viscoelastic model (SDVES) which addresses some of the shortcomings of the traditional models. SDVES corrects the single timescale, viscous dissipation term with a non-linear strain dependent prefactor. We see that the SDVES model provides a significant improvement in capturing experimental observations, such as the nonlinear increase in force at linear application of strain, the decay of force to a non-zero force at static strain application, the absence of a tensile load at the end of decompression and the recovery of peak forces during cyclic loading after waiting for a time $t>\tau$ at $\epsilon=0$. The SDVES model is effective in this aspect for both strain and stress driven boundary conditions, in two chemically entirely different systems, indicating the robustness of the non-linear correction. The proposed model works optimally at low strain rates. At higher strain rates we see certain inconsistencies that suggest that future work should focus on refining the model rate dependence in the model by including multiple rate constants or additional viscous mechanisms.




\nocite{*}

\acknowledgements

We thank the Lorentz Center Workshop ``Getting Into Shape'' for stimulating the discussions that lead to this manuscript. This project has received funding from the European Union’s Horizon 2020 research and innovation program under the Marie Skłodowska Curie grant agreement number 812638.
This work benefited from the technical contribution of the joint service unit CNRS UAR 3750. The authors would like to thank the engineers of this unit for their advice during the development of the experiments.

\appendix

\section{\label{tradmodel} Traditionally used mechanical models}

In a viscoelastic model the combination between elastic reversible response and viscous irreversible one is addressed in a constitutive way, i.e. in a way that depends upon the material under investigation.
Rheologically, such models are represented by a combination of springs and dashpots.
We noted that spring is an element reacting with a force $F_s$ that is proportional to the relative displacement $u_s$ between the two ends of a spring,
\begin{equation}
    F_s= k_{el} u_s,
\end{equation}
where $k_{el}$ is the stiffness of the spring. 
In a similar consideration, dashpot is an element reacting with a force $F_d$ that is proportional to the relative velocity $v_d$,
\begin{equation}
   v_d = \frac{\partial u_d}{\partial t}=\dot{u}_d, \qquad F_d= c v_d,
\end{equation}
where $u_d$ is the relative displacement between the two ends of a dashpot, the superimposed dot is the time derivative operator and $c$ is the viscosity coefficient. We use here displacements instead of the usual strain to more easily make a clear comparison with experimental data.

\subsection{Kelvin–Voigt (KV) model}
In a Kelvin–Voigt model the combination between elastic and viscous response is the parallel of a spring and a dashpot. 
The reaction force $F_{KV}$ is the sum of that of the spring, e.g. with stiffness $k_{KV}$, and of the dashpot, e.g. with viscosity $c_{KV}$,
\begin{equation}\label{kvmodel}
    F_{KV}=k_{KV} u_{KV}+ c_{KV} \dot{u}_{KV}.
\end{equation}

The KV model used a spring linear with displacement with a dashpot linear with the velocity thus it cannot recreate the nonlinear evolution of force with a constant velocity as shown in figure \ref{fig:Kelvin Voigt}. In protocol 1, the $\dot{u}_{KV}=0$, when no load is applied and $\dot{u}_{KV}=constant$ when compression begins, this the force response jumps from zero to a finite value even at small $u_{KV}$. During relaxation, $\dot{u}_{KV}=0$, thus the force response suddenly drops to only the elastic response. And during decompression, $\dot{u}_{KV}=-constant$, so towards the end of the decompression the negative viscous force overpowers the small positive elastic response resulting in a net adhesive force that do not match the experimental results.

\subsection{Maxwell (M) model}

Another model that is often used in viscoelasticity is the Maxwell model. 
Here, the combination between elastic and viscous response is the series of a spring and a dashpot. Thus, the relative displacement $u_M$ between the two sides of the Maxwell element is the sum between those of the spring $u_{Ms}$ and of the dashpot $u_{Md}$.
\begin{equation}\label{m1}
    u_M=u_{Ms}+u_{Md}.
\end{equation}
Besides, the reaction force $F_M$ is common to the two elements in series,
\begin{equation}\label{m2}
    F_M=k_M u_{Ms}=c_M \dot{u}_{Md} 
\end{equation}
where $k_M$ and $c_M$ are, respectively, the stiffness of the spring and the viscosity of the dashpot of the Maxwell system.
However, the only measurable quantity is the relative displacement $u_M$ defined in Eq.~\ref{m1}. 
Thus, let us derive with respect to time the first of Eq.~{m2} and insert the Eq.~{m1},
\begin{equation}\label{m3}
    \dot{F}_M=k_M \left( \dot{u}_M-\dot{u}_{Md}\right).
\end{equation}
We obtain, by insertion into Eq.~{m3} of the second of Eq.~{m2}, a differential equation
\begin{equation}\label{m4}
    \dot{F}_M + \frac{k_M}{c_M} F_M=k_{M} \dot{u}_M,
\end{equation}
that can be analytically solved once an initial condition $F_M(t=0)=0$ on the reaction force is prescribed. 
The solution is given in the following integral form,
\begin{equation} \label{m5}
    F_M (t)=\int_0^t k_M \dot{u}_M (\alpha)  \exp\left[\frac{k_M}{c_M} \left( \alpha -t \right) \right] d \alpha.
\end{equation}
Eq. \ref{m5} gives the reaction force as a function of the history of the imposed velocity $\dot{u}_M$ with a kernel $\exp\left[k_M \left( \alpha -t \right) /c_M \right]$ that prescribes the weight of such a history. 
In particular, the value $\dot{u}_M(\alpha=t)$ of the imposed velocity at the present time $\alpha=t$ is weighted more than that $\dot{u}(\alpha=0)$ at the beginning of the deformation history (when the initial condition is prescribed) because of the structure of the exponential kernel and of the positiveness of both $k_M$ and $c_M$.
It is worth noting, from the expression Eq.~\ref{m5}, that the ratio $c_M/k_M$  gives us the characteristic time of the Maxwell system, which we shall call $\tau_M$.

However, even the Maxwell model is not sufficient to catch the experimental results. 
If an asymptotically constant displacement, i.e. $u_M=const$, is imposed for values of time $t>>t_M+c_M/k_M$, then the solution (Eq.~{m5}) prescribes asymptotically, i.e. for time $t>>t_M+c_M/k_M$, a negligible small reaction. 
However, if we look at the results shown in figure \ref{fig:relaxation_diff_protocol}d we clearly see that, at a finite applied strain, the material's response asymptotically tend to a finite non-zero and positive value. Thus, in order to overcome this problem, we could add another spring in parallel to obtain the standard model presented in the next subsection.

\subsection{Standard Linear Solid(SLS) model}
Another model that is often used in viscoelasticity is the standard linear solid model.
Here, the combination between elastic and viscous response is the parallel of a spring and a Maxwell element. 
Thus, the relative displacement $u_{st}$ between the two sides of the standard element is the same for the spring, e.g. with stiffness $k_{sts}$, and for the Maxwell element, e.g. with stiffness $k_{stM}$ and viscosity $c_{stM}$.
Its reaction force $F_{SLS}$ is the sum of that of the spring $F_{sts}$ and of the Maxwell element $F_{stM}$,
\begin{equation}\label{st1}
    F_{SLS}=k_{sts} u_{st}+ F_{stM},
\end{equation}
where the reaction force of the Maxwell system is the same of that of its constituting spring and dashpot, i.e.,
\begin{equation}\label{st2}
    F_{stM}=k_{stM} u_{stMs}=c_{stM} \dot{u}_{stMd},
\end{equation}
where the relative displacement between the two sides of the Maxwell element $u_{st}$ is the sum of those between the spring $u_{stMs}$ and the dashpot $u_{stMd}$, belonging both to the Maxwell element of the standard model,
\begin{equation}\label{st3}
    u_{st}=u_{stMs}+u_{stMd}.
\end{equation}
Differentiating  Eq.~{st1} with time and inserting $\dot F_{stM}$ after differentiating Eq.~\ref{st2} with time we get:
\begin{equation}\label{st4}
    \dot{F}_{st}=k_{sts} \dot{u}_{st}+ k_{stM} \dot{u}_{stMs}.
\end{equation}
Insertion of (\ref{st3}) into (\ref{st4}) yields
\begin{equation}\label{st5}
    \dot{F}_{st}=k_{sts} \dot{u}_{st}+ k_{stM} \left(\dot{u}_{st}-\dot{u}_{stMd}\right).
\end{equation}
and use of the second of (\ref{st2}) and of (\ref{st1}) into (\ref{st5}) yields,
\begin{equation}\label{st6}
    \dot{F}_{st}=k_{sts} \dot{u}_{st}+ k_{stm} \left(\dot{u}_{st}-\frac{F_{st}-k_{sts} u_{st}}{c_{stm}}\right).
\end{equation}
Eq. \ref{st6} is the differential equation of the standard linear solid model, that can be analytically solved as follows in an integral form,
\begin{eqnarray} \label{st7}
   F_{st} (t)= \frac{1}{c_{stM}}\int_{0}^{t}\Biggr[ 
   \exp\biggl(\frac{k_{stM}}{c_{stM}}\left(\alpha-t\right)\biggl) 
   \\
   \times\{k_{stM}k_{sts}u_{st}\left(\alpha\right)+c_{stM}\left(k_{stM}+k_{sts}\right)\dot{u}_{st}\left(\alpha\right)\} \Biggr] d \alpha. \nonumber
\end{eqnarray}
where, the trivial initial condition $F_{st}(t=0)=0$ for the reaction force is also prescribed.
Eq. \ref{st7} gives the reaction force for the standard model as a function of the history of the imposed displacement and velocity with a kernel (similar to that of the solution of the Maxwell model in Eq.~\ref{m5}), that prescribes the weight of such a history. 
In particular, the values $u(\alpha=t)$ and $\dot{u}(\alpha=t)$ of the imposed displacement and velocity at the present time $\alpha=t$ is weighted more than that $\dot{u}(\alpha=0)$ and $\dot{u}(\alpha=0)$ at the beginning (when the initial condition is imposed) because of the structure of the exponential kernel $\exp\left[k_{stM}\left(\alpha-t\right)/c_{stM}\right]$ and of the positiveness of both $k_{stM}$ and $c_{stM}$.

On the one hand, the presence of the first addend of the second line of Eq.~\ref{st7} solves the problem mentioned at the previous subsection and the response to a prolungated constant non-null displacement is non-null.
On the other hand, this standard linear solid model formulation is still not sufficient to catch the phenomenology that is here investigated.
The first reason is that for an imposed displacement $u_{st}$ that is initially positive and asymptotically null, i.e., $u_{st}=0$ for values of time $t>t_{st}$ larger than a certain value $t_{st}$ and therefore for negative velocity before (i.e., for $t<t_{st}$) that value, the solution Eq.~\ref{st7} prescribes, for time $t>t_{st}$, negative reaction force due to the mentioned negative sign of the imposed velocity for time $t<t_{st}$, and this is not convergent with the experimental evidence as can be seen from figures \ref{fig:tradmodel}.
The second reason is that for an imposed displacement that is asymptotically constant, i.e. $u_{st}=\bar{u}_{st}$ for values of time $t>t_{st}$ larger than a certain value $t_{st}$, the solution Eq.~\ref{st7} prescribes asymptotically, i.e., for time $t>>t_{st}+c_{stM}/k_{stM}$, a reaction force that is linear with respect to $\bar{u}_{st}$ and proportional to $k_{sts}$.
This means that, asymptotically, the reaction is driven by the linear spring that is, in the standard model, in parallel with the Maxwell system. 
However, experimental evidence prescribes nonlinear elastic behavior. 
Thus, we formulated a new model that is coherent with the experimental evidence in Sec.~\ref{new_model}.

\section{\label{para_meters}Experimental and Model settings}
This appendix lists the fit parameters used with the SDVES model for the different figures generated are listed in table \ref{tab:parameters}. All models use $\tau=1500 s$ to predict the response of the 3D packing. The parameter table also suggests that there is a relation $a(\dot\epsilon)$. We attribute this rate dependence to other time scales that may be present in the microscopics of the particles, as discussed in the main text.

\begin{table}[h]
\caption{\label{parameter_experiment}Experimental parameters for the protocols listed in Section~\ref{exp_protocols}. In Protocol 6, we used a range of strain rates.}
\begin{tabular}{|l|l|l|l|l|}
\hline
Protocol & sample height ($l$) & $\dot\epsilon$                          & $t_{wait}$ & $\epsilon_{max}$ \\
         & {[}mm{]}        & $[s^{-1}]$       & $[s]$             & \%          \\ \hline
1        & 88.15           & $3\times 10^{-5}$                     & 1000                & 3           \\
2        & 91.2            & $2\times 10^{-3}$                     & 3000                & 3.5 - 4     \\
3        & 90.3            & $2\times 10^{-3}$                     & 0                   & 3.5 - 4     \\
4        & 90.3            & $2\times 10^{-3}$                     & 0                   & 5.5         \\
5        & 88              & $2\times 10^{-3}$                     & 0                   & 0.5 -5.5    \\
6        & 88.15           & $3\times10^{-5}$ \ldots   & 1000                & 3   \\
        &            &$1.67\times10^{-2}$ &                 &    
\\ \hline
\end{tabular}
\end{table}

\begin{table}[h]
    \begin{center}
    \caption{\label{parameter_table}Model Parameters for SDVES model}
     \begin{tabular}{| l | c | c | c | c | c | c | c | c | c |}
     \hline
            \rotatebox{90}{Protocol} & \rotatebox{90}{Figure} & \rotatebox{90}{$a$} & \rotatebox{90}{$k_N~[N/m^{a}]$} & \rotatebox{90}{$c_N~[Ns/m]$} &  \rotatebox{90}{$\tau ~[s]$} & \rotatebox{90}{$w~[-]$} & \rotatebox{90}{$b~[-]$}& \rotatebox{90}{$u_c~[mm]$}& \rotatebox{90}{$\dot\epsilon [s^{-1}]$}\\
            \hline
            1 & \ref{good_fits}a & 4 &$2.73G$  & 410 & $1500$ &6 &2 &$2.6 $&$2\times10^{-5}$\\
            1 & \ref{good_fits}a & 3 &$5.5M$  & 700 & $1500$ &6 &2 &$2.6 $&$2\times10^{-5}$\\
            2 & \ref{good_fits}b & 3 &$5M$ & 700 & $1500$ &6 &2 &$3.8 $&$2\times10^{-3}$\\
           3 & \ref{good_fits}c & 2.5 &$0.39M$ & 900 & $1500$ &6 &2 &$5.3 $&$2\times10^{-3}$\\
           6 & \ref{good_fits}d & 4 &$2.73G$  & 410 & $1500$ &6 &2 &$2.6 $&$2\times10^{-2}$\\
           6 & \ref{good_fits}e,f & 4 &$2.73G$  & 410 & $1500$ &6 &2 &$2.6 $&variable\\
           3 & \ref{good_fits}g & 4 &$1.36M$ & 400 & $1500$ &6 &2 &$3.6 $&$2\times10^{-3}$\\
           4 & \ref{good_fits}h & 2.5 &$0.39G$ & 900 & $1500$ &6 &2 &$5.3 $&$2\times10^{-3}$\\
           5 & \ref{good_fits}i & 2.5 &$0.39G$  & 900 & $1500$ &6 &2 &$5.3 $&$2\times10^{-3}$\\
           x & 6a & 2.5 & $1.65G$ & 50k & $50$ & 4 & 2.5 & 0.15 &variable\\
           x & 6d & 2.5 & $18M$ & $20k$ & $50$ & 4 & 2.5 & 1.2 &variable\\
           
   \hline     
    \end{tabular}
    \label{tab:parameters}
    \end{center}
\end{table}
\clearpage
\bibliography{apssamp}

\end{document}